\documentclass[12pt, epsf]{JHEP}

\usepackage{epsfig}

\def\pic #1#2{\hbox{\lower#1pt\hbox{\mbox{\epsfxsize=16truemm 
\epsffile{#2}}}}}
\def\picl #1#2{\hbox{\lower#1pt\hbox{\mbox{\epsfxsize=30truemm 
\epsffile{#2}}}}}

\def\picel #1#2{\hbox{\lower#1pt\hbox{\mbox{\epsfxsize=50truemm 
\epsffile{#2}}}}}

\newcommand{\non}{\nonumber \\*}

\newcommand{\eq}[1]{Eq.~(\ref{#1})}
\def \ov {\over }
\def\bea{\begin{eqnarray}}
\def\eea{\end{eqnarray}}

\def\LB{\left(}
\def\RB{\right)}
\def\be{\begin{equation}}
\def\ee{\end{equation}}
\def\la{\label}
\def\e{\epsilon}
\def\a{\alpha}
\def\b{\beta}
\def\A{{\cal A}}
\def\ex{{\rm exp}}

\def\B{{\cal B}}

\font\mybb=msbm10 at 12pt
\def\bb#1{\hbox{\mybb#1}}
\def\re {\bb{R}}

\font\mybbb=msbm10 at 20pt
\def\bbb#1{\hbox{\mybbb#1}}
\def\Ree {\bbb{R}}

\font\mycc=msbm10 at 14pt
\def\cc#1{\hbox{\mycc#1}}
\def\rr {\cc{R}}

\def\x{{\hat x}}
\def \tr {{\rm tr}}
\def \de{\partial}
\def\R{{\cal R}}
\def\det{{\rm det}}
\def\T{{\cal T}}
\def\l{{\cal L}}
\def\E{{\cal E}}
\def\V{{\cal V}}
\def\S{{\cal S}}

\def\C{{\cal C}}
\def \bi{\bibitem}
\def\I{{\cal I}}
\def\g{\tilde g}

\title{\LARGE Renormalization of  Quantum Field Theories on 
Noncommutative $\Ree^d$, I. Scalars}
\author{Iouri Chepelev and Radu Roiban\\C.N. Yang Institute for Theoretical Physics\\
SUNY at Stony Brook NY11794-3840, USA\\Email: chepelev@insti.physics.sunysb.edu and 
roiban@insti.physics.sunysb.edu}

\abstract{
A noncommutative Feynman graph   is
a ribbon graph and  can be drawn on a genus $g$ 2-surface with
a boundary. 
We formulate a general convergence theorem for the noncommutative 
Feynman graphs in topological terms and prove it for some classes of 
diagrams 
in the scalar field theories. We propose 
a noncommutative analog of Bogoliubov-Parasiuk's recursive 
subtraction formula and show 
that the subtracted graphs from a class $\Omega_d$ satisfy the 
conditions of the 
convergence theorem.  For a generic
 scalar noncommutative quantum field theory 
on $\re^d$, the class $\Omega_d$ is  smaller than the class of all 
diagrams in the theory. 
This leaves open the  question of perturbative renormalizability
of noncommutative field theories. 
We comment on how the supersymmetry can improve the situation
and  suggest  that  a  noncommutative  analog of Wess-Zumino model
is renormalizable.
}
\preprint{ITP-SB-99-61}

\begin{document}

\section{Introduction}
\subsection{Historical background}
What would physics  be like if the space in which it took place was
not a set of points, but a non-commutative space\footnote{
For a comprehensive account of noncommutative geometry, see
ref.\cite{conbook}.}? This was the question asked by Connes in
ref.\cite{connes} where it was  shown that a small 
modification of the usual picture of space-time gives an alternative 
explanation of the  Higgs fields and of the  way they appear in the
Weinberg-Salam model\footnote{See also \cite{conlot}}. Field theories on noncommutative spaces (NFT) \cite{filk,chaichian}
are also interesting as a first step towards a formulation of 
quantum gravity which avoids standard problems\cite{dfr}.

NFT became 
popular in the community of string theorists with the appearance of 
a paper by Connes, Douglas and Schwarz \cite{cds97}, where it was 
argued that M-theory in a 
constant  three-form tensor background is equivalent to  a super Yang-Mills
theory on a noncommutative torus. For a review  of 
developments following 
 ref. \cite{cds97}, see ref.\cite{dougrev}. 
A second wave of interest towards NFT came with
the work of Seiberg and Witten \cite{sw99} which summarized and 
extended  earlier ideas 
about the appearance of noncommutative geometry in string theory with
a nonzero B-field\footnote{An extensive list of references on the
subject can be found in ref.\cite{sw99}}. 

As stressed in ref.\cite{dougrev}, the most pressing question regarding NFT
is whether or not the quantum theory (NQFT) is well-defined. 
The algebra of functions on the noncommutative $\re^{d}$ is isomorphic
to the algebra of functions on commutative $\re^{d}$ with
the multiplication of functions given by the 
 $\star$-product 
\be
(\phi_1\star \phi_2)(x)=
 \left. {\rm e}^{i \theta_{\mu\nu}{\de\ov \de\xi^{\mu}}{\de\ov \de\zeta^{\nu}}}
\phi_1(x+\xi) \phi_2(x+\zeta)\right|_{\xi=\zeta=0}
\ee
The NFT action is the usual field theory action where the point-wise
multiplication of the fields is replaced by the $\star$-product. The  
non-locality of the NFT action in the position space looks  bad 
at first sight and  one might be led to conclude that NQFT 
 is  perturbatively non-renormalizable. It was  pointed
out in ref.\cite{dougrev} that after deriving the Feynman rules for
NQFT and studying the one loop amplitude in momentum space 
one sees that the situation is actually rather good because the 
non-local interaction terms
in the action provide oscillatory factors in the Feynman integrals.
Indeed, one-loop
renormalizability of noncommutative Yang-Mills (NYM) theory has 
been demonstrated in 
ref.\cite{martin}. In  ref.\cite{lattice}  
 a noncommutative version of Wilson's lattice gauge theory
formalism was developed. Such a formalism has the potential of 
clarifying issues of renormalization.

In ref.\cite{filk}, Filk analyzed  the structure of  
Feynman diagrams for the NQFT. He pointed out that 
the planar diagrams do not have oscillatory factors (involving loop momenta)
 coming from
the non-local interaction terms, and thus the corresponding integrals 
are  the same as in usual QFT. 

The NQFT and QFT amplitudes for
a planar graph $G$ are related as
\be
I_{\rm  NQFT}(G,k)=e^{i\varphi(k)}I_{\rm  QFT}(G,k)
\la{varphi}
\ee
where $k$ denotes the external momenta and $\varphi(k)$ is a phase depending
only on $k$.

This means that  the planar
diagrams of NQFT  diverge in  the same way
as the corresponding QFT diagrams. On the other hand, all non-planar
diagrams 
 have the oscillatory factors involving loop momenta.
In
ref.\cite{susskind}, Bigatti and Susskind claimed that the oscillatory factors would regulate divergent 
diagrams and make them finite, unless the diagrams contained 
divergent planar subdiagrams.\footnote{The claim made in 
ref.\cite{susskind} was partially 
supported by the supergravity calculations of gauge-invariant quantities
of large-N noncommutative SYM  in ref.\cite{russo}. See also 
ref.\cite{hashimoto}}

\subsection{Logic and structure of the paper}
It is  a desirable property
of a diagram that it diverges only when it contains divergent 
planar subgraphs.  The reason is the following. Let G be a non-planar 
NQFT graph 
which does not contain divergent planar subgraphs. If  $G$ were  divergent, it
would  have to diverge {\it properly}, i.e. it should be possible to
subtract the divergences by the introduction of  counterterms
which have the same form as those already occurring in the action. 
It is very unlikely that the divergent part of an integral 
involving
oscillatory functions  is proportional to the phase factors appearing in the
Lagrangian.\footnote{In the minimal subtraction approach of ref.\cite{martin}
there are some unusual divergent terms coming from the integrals 
involving periodic functions, but these terms cancel in the sum over 
all one-loop  diagrams. The underlying reason for such  cancellations
seems to be the convergence of non-planar diagrams.}

In this paper we will analyze  scalar field theories on noncommutative
$\re^d$.\footnote{Yang-Mills theory will be analyzed in
ref.\cite{YMchep}.} 
Our analysis consists of four steps:
\begin{enumerate}
\item A formulation of the convergence theorem for noncommutative Feynman graphs.
\item A recursion formula for the subtraction of divergences.
\item A proof that the application of the recursion formula  
to the integrand of a noncommutative graph yields an expression satisfying 
the conditions of the convergence theorem.
\item A proof  that the subtraction 
procedure is equivalent to the introduction of the counterterms  which have the same form as those already occurring in the action. 
\end{enumerate}   

\FIGURE[ht]{
\includegraphics[width=7truecm]{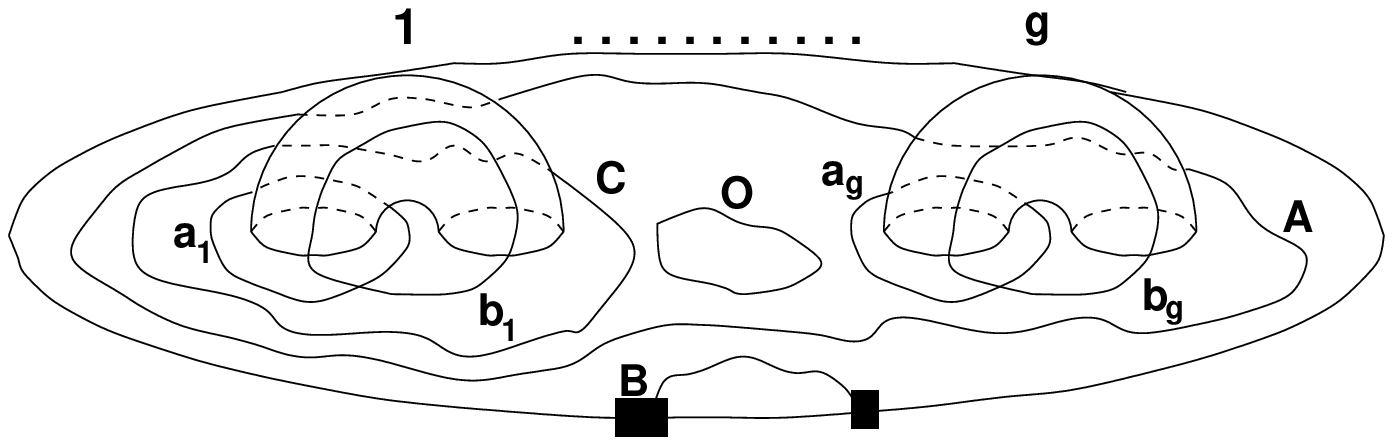}
\hspace{.5truecm}
\includegraphics[width=7truecm]{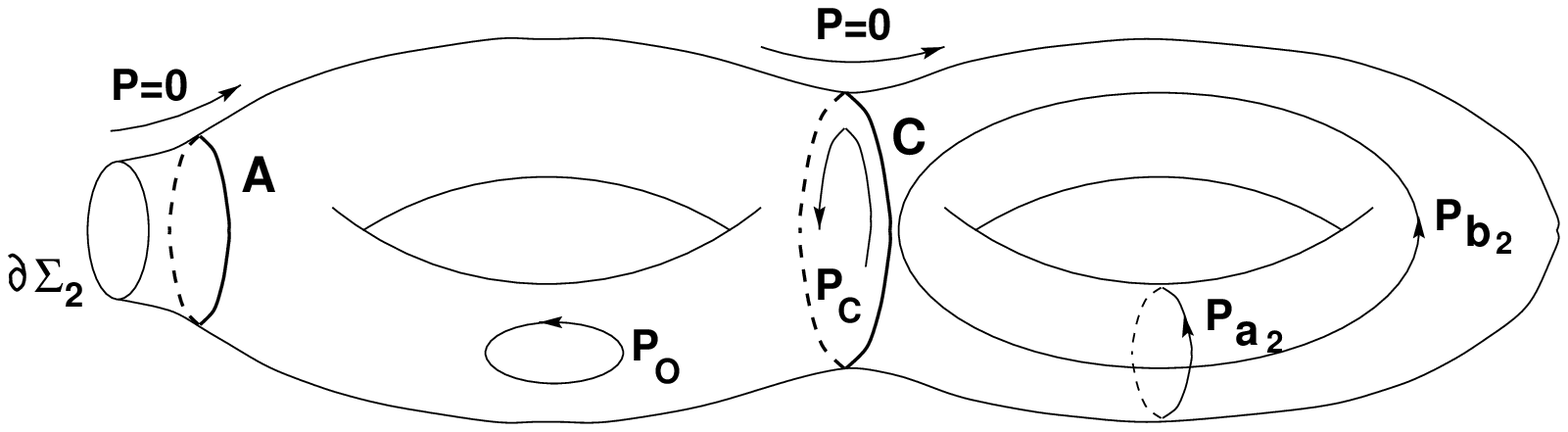}
{\vspace{2truemm}}
\centerline{~~{\bf (a)}A genus $g$ surface with a boundary
~~~{\bf (b)}Momentum flow on a $g=2$ surface $\Sigma_2$}
{\vspace{-8truemm}}
\caption{}
}


The step 1  (the convergence theorem) is 
central to the analysis.
We will find  precise conditions under which the claim made 
in ref.\cite{susskind} regarding the
convergence of noncommutative graphs is  realized. 
A general NQFT Feynman graph with some external lines can be drawn on
a genus $g$ surface with a boundary (with one end of each external line 
being attached to the boundary).   
Let G be a NQFT Feynman
graph. Draw it on a 2-surface  $\Sigma_g$ of genus $g$ with a boundary 
$\partial \Sigma_g$. The non-trivial cycles of $\Sigma_g$ are
 $a_1,b_1,\ldots,
a_g,b_g$ (see figure 1).  Cycles $A$, $B$, $C$ and $0$ are trivial.\footnote{A cycle on $\Sigma_g$ is called non-trivial if it is
a non-trivial element of the first homology group $H_1(\Sigma_g)$.
In addition to the  trivial cycles that are contractible to a point,
there are trivial cycles which are not contractible to a point.   
For example, cycles $A$ and $C$  in figure 1 are trivial because 
$A=a_1b_1a_1^{-1}b_1^{-1}\cdots a_gb_ga_g^{-1}b_g^{-1}$ and 
$C=a_1 b_1 a_1^{-1} b_1^{-1}$, i.e. $A$ and $C$ are commutants. See
ref.\cite{springer} for the details.}
Let $\gamma$ be 
a subgraph of $G$. Let $c(\gamma)$ be the number of
inequivalent  non-trivial cycles of $\Sigma_g$  spanned by the 
closed paths in $\gamma$.
 To the subgraph $\gamma$ we assign an index $j(\gamma)=0 {\rm ~or~}1$ 
which characterizes the non-planarity of $\gamma$ with respect to
the external lines of $G$.\footnote{For the precise definition of $j(\gamma)$
see section 3.}

\noindent
{\it Example 1.} A noncommutative Feynman graph in figure 2(a) is shown in 
figure 2(b) 
as a ribbon graph on a genus one 2-surface with a boundary. In $d$ dimensions
$\omega(\gamma)$, $c(\gamma)$ and $j(\gamma)$ for some 
subgraphs $\gamma$ read as follows. 
$$
\begin{array}{cccc}
\gamma&\omega(\gamma)&c(\gamma)&j(\gamma) \\
(239)&d-6&0&0 \\
(45679)&d-10&1&1\\
(2345678)&2d-14&2&1
\end{array}
$$

Our convergence theorem can be stated as follows.
\noindent
{\it  The 1PI graph $G$ 
is convergent  if and only if for any subgraph $\gamma\subseteq G$ 
at least one of the following conditions is satisfied:(1)
$\omega(\gamma)-c(\gamma)d<0$, (2) $j(\gamma)=1$}.\footnote{For
the scalar field theories discussed in this paper,  
$\omega(\gamma)=d L(\gamma)-2I(\gamma)$, where $L$ and $I$ are the number of 
independent  loops and internal lines of $\gamma$ respectively. It is 
assumed that the external momenta of the graph G are non-exceptional.} 

The meaning of this convergence theorem is the following.
Each handle in figure 1 has two nontrivial cycles $a_i$ and $b_i$. Let 
$p_{a_i}$ and 
$p_{b_i}$ 
be the total internal momenta flowing through the graph $\gamma$ 
along the cycles $a_i$ and $b_i$ respectively. 
The phase factor associated with each handle is
$\ex (i\theta_{\mu\nu}p_{a_i}^{\mu}p_{b_i}^{\nu})$. As far as the 
convergence property of the graph is concerned, the effect of 
this phase factor is equivalent to reducing the number of loops
by two. The condition $j(\gamma)=1$ for
a subgraph $\gamma\subset G$ means that a certain combination $\sum P_v$ 
of external momenta
of $G$ flows through $\gamma$ and the path of the flow is 
{\it not homologous} to cycle $B$. The phase factor 
associated with such a flow is, schematically,
$\ex (i\theta_{\mu\nu}(\sum P_v)^{\mu} q^{\nu})$, where $q$ is the 
loop momentum
along a combination of cycles $a_1,\ldots,b_g$. This  phase
factor makes $\gamma$ finite for arbitrary $\omega(\gamma)$ because of the
exponential suppression at large external momenta (see Section 3 for details).

\FIGURE[ht]{
\includegraphics[width=4truecm]{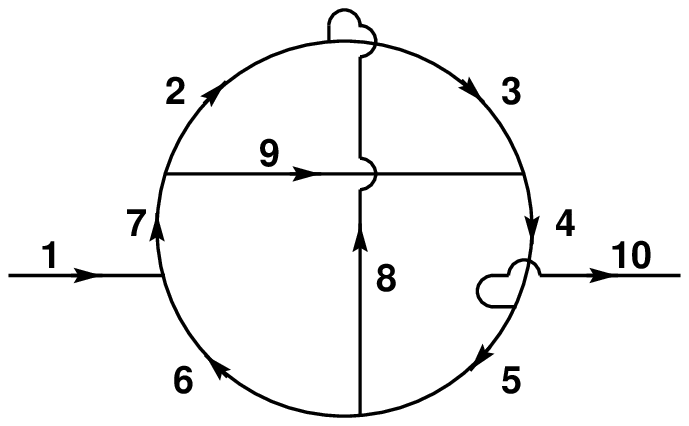}
\hspace{1truecm}
\includegraphics[width=8truecm]{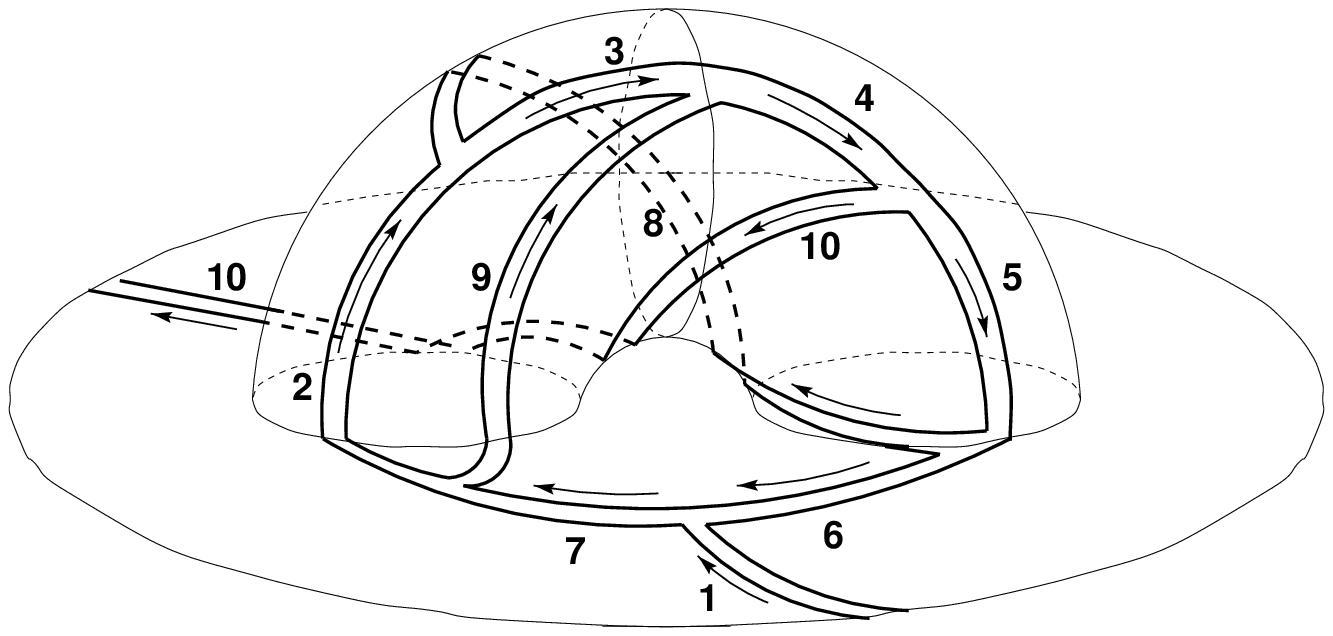}
{\vspace{2truemm}}
\centerline{~~~~~~~{\bf (a)}~~~~~~~~~~~~~~~~~~~~~~~~~~~~~~~~~~~~~~~~~~~~~~~~~~{\bf (b)}~~~~~~~~~~~~~~~~}
{\vspace{-8truemm}}
\caption{Illustration for example 1.}
}


The steps 2,3 and 4  of our analysis are  straightforward generalizations
of the corresponding steps in the proof of renormalizability of 
commutative QFT. The only complication that arises due to the 
noncommutativity of the space is the distinction between topologically
trivial and nontrivial subgraphs. For a given NQFT in $d$ dimensions,
we will show that the Feynman integral  for any graph in
a class $\Omega_d$ (to be defined in section 4)  can be made finite by 
the application of the recursive subtraction
formula to the integrand of that integral.

The paper is organized as follows. In section 2 we review Feynman
rules for scalar NQFT and derive parametric integral 
representation for the amplitudes.
We begin section 3 by analyzing the convergence properties of
some simple diagrams and demonstrating the convergence
of some classes of diagrams.
We  formulate a general convergence theorem, and show how it
{\it explains}, in a unified manner, the convergence of the diagrams
analyzed earlier. 
In section 4 we write down an analog of Bogoliubov-Parasiuk's recursion
formula for the subtraction of divergences in NQFT and  
prove the  convergence of  subtracted integrals for the graphs from
class $\Omega_d$.
We then suggest that the supersymmetric extension of scalar NQFT  is
renormalizable.

\section{Scalar NQFT on the noncommutative $\rr^d$}

\subsection{Definition of  NQFT and Feynman rules}
The noncommutative $\re^d$ is defined as follows.
The coordinates  $x^{\mu}$($\mu=1,\ldots,d$) of commutative $\re^d$ are 
replaced by 
the self-adjoint operators $\x^{\mu}$ in a Hilbert space ${\cal H}$ satisfying
the commutation relations 
\be
[\x^{\mu},\x^{\nu}]=2i \theta_{\mu\nu},\,~~~~~ [\theta_{\mu\nu},\x^{\rho}]=0
\ee
where $\theta$ is a non-degenerate $d\times d$ skew-symmetric matrix ($d$ is
even).

With a function $\phi(x)$ on the commutative space $\re^d$ one associates
the operator $\Phi(\x)$ acting in the Hilbert space ${\cal H}$ using
the rule:  
\be
\Phi(\x)={1\ov (2\pi)^d} \int d^dx d^dk  {\rm e}^{ik_{\mu}(\x^{\mu}- 
x^{\mu})}\phi(x)
\ee
Given an operator $\Phi(\x)$, the function $\phi(x)$ can be obtained using
\be
\phi(x)={1\ov (2\pi)^{d/2}} \int d^dk {\rm e}^{i k_{\mu}x^{\mu}} \tr 
\Phi(\x) {\rm e}^{-i k_{\mu} \x^{\mu}}
\la{phivbphi}
\ee
where the trace $\tr$ is over the Hilbert space ${\cal H}$.

To the product of two operators $\Phi_1$ and $\Phi_2$ corresponds  a
 $\star$-product
\bea
(\phi_1\star \phi_2)(x)&=&{1\ov (2\pi)^{d/2}}\int d^dk {\rm e}^{ik_{\mu}x^{\mu}}
\tr[\Phi_1\Phi_2 {\rm e}^{-i k_{\mu}\x^{\mu}}] \non
&=& \left. {\rm e}^{i \theta_{\mu\nu}{\de\ov \de\xi^{\mu}}{\de\ov \de\zeta^{\nu}}}
\phi_1(x+\xi) \phi_2(x+\zeta)\right|_{\xi=\zeta=0}
\eea

The noncommutative  analog of the classical massive scalar field theory action 
(without derivative couplings) on a commutative space reads  
\be
{\tilde S}[\Phi] =\tr \LB \sum_{\mu} ({1\ov 2}\theta^{-1}_{\mu\nu}[\x^{\nu},\Phi(\x)])^2+ {m^2\ov 2} \Phi(\x)^2 + {g\ov n} (\Phi(\x))^n \RB 
\ee
This action can be expressed in terms of $\phi(x)$ defined by \eq{phivbphi} as
\be
S[\phi]=\int d^d x [{1\ov 2}(\partial_{\mu}\phi)^2+{m^2\ov 2} \phi^2 +
{g\ov n}~(\phi\star\cdots\star\phi) ]
\la{nqft}
\ee
The action \eq{nqft} in the  momentum space reads  
\be
S[\phi]=\int d^dk {1\ov 2}
\phi(-k)(k^2+m^2)\phi(k)+\int d^d k_1 \cdots d^dk_n 
V(k_1,\ldots,k_n) \phi(k_1)\cdots\phi(k_n)
\ee
where
\be
V(k_1,\ldots,k_n)=
{1\ov n}\delta(k_1+\ldots+k_n) {\rm exp}\LB\sum_{i<j}
k_i^{\mu}k_j^{\nu}\theta_{\mu\nu}\RB
\la{phase}
\ee

\FIGURE[ht]{
\includegraphics[width=9truecm]{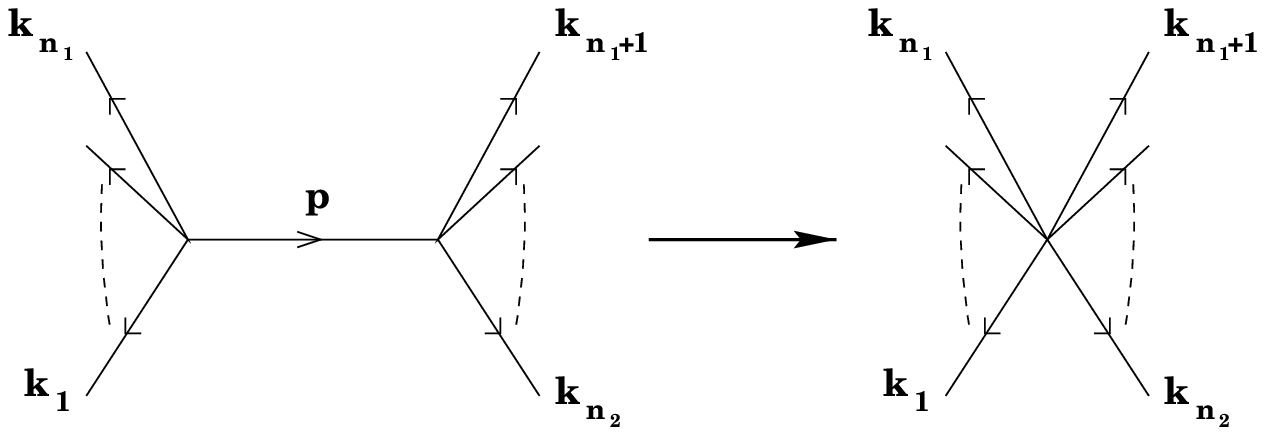}
\caption{Filk's operation 1.}
}


\FIGURE[ht]{
\includegraphics[width=9truecm]{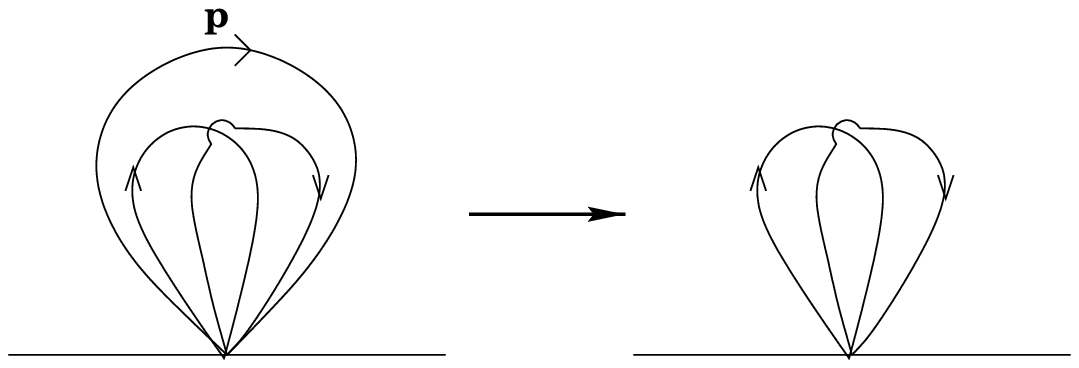}
\caption{Filk's operation 2.}
}


Due to the noncommutativity of the $\star$-product, the interaction
term in $S[\phi]$ is not totally symmetric under the exchange of the arguments,
but only under  cyclic permutations. This implies that the 
Feynman graphs of NQFT are equivalent to ribbon graphs. Thus, a
general diagram of NQFT can be drawn on a genus $g$ 2-surface. 
A NQFT Feynman graph  may have crossings of  internal and external
lines.  In order to find the contribution of the phase factors \eq{phase} to
an arbitrary Feynman graph $G$, one may find it useful  first to apply 
the following 
operations to $G$ \cite{filk} (the same set of operations were first
introduced in a different context in ref.\cite{nakayama}):

\noindent
(1)Contraction of two vertices connected by a line  (figure 3):
$$
V(k_1,\ldots,k_{n_1},p)V(-p,k_{n_1+1},\ldots,k_{n_2})=
V(k_1,\ldots,k_{n_2})\delta(k_1+\ldots+k_{n_1}+p)
$$
\noindent
(2)Elimination of a loop which does not cross other lines (figure 4):
$$
V(k_1,\ldots,k_{n_1},p,k_{n_1+1},\ldots,k_{n_2},-p)=
V(k_1,\ldots,k_{n_1},k_{n_1+1},\ldots,k_{n_2})  {\rm ~if} 
\sum_{i=n_1+1}^{n_2} k_i=0
$$
These operations are based on momentum conservation 
and cyclic symmetry at each vertex. Using these  two operations one
may reduce any Feynman graph to a graph which consists of only one vertex. 

For a planar graph this reduction leads to a one vertex graph with
external lines. For a non-planar graph this reduction leads to
a rosette\footnote{Different reductions may give different rosettes, 
but all of them give the same phase factor.}. 
The set of rosette lines of a graph $G$ is
denoted as $\R(G)$. 

\FIGURE[ht]{
\includegraphics[width=4truecm]{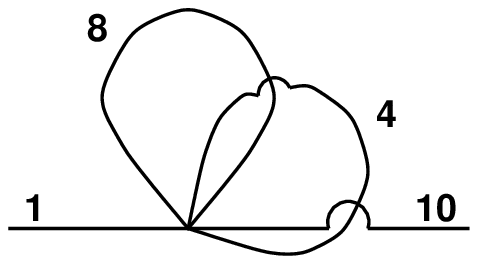}
\caption{Illustration for example 2.}
}


\noindent
{\it Example 2.}
Shrinking lines 2,7,6,5 and then 3,9  in figure 2(a), one finds the rosette
of figure 5. In this case $\R=\{ 4,8\}$.

\noindent
Let us define $I_{ij}(G)$ to be the intersection matrix
of internal lines of an oriented graph $G$ 
(orientation is given by the sign convention chosen for the momenta in the conservation conditions):
If  $i,j\in \R(G)$, then
\be
I_{ij}(G)=\left\{
\begin{array}{cl}
1&{\rm line~}j{\rm ~crosses~}i{\rm ~from~right}\\ 
-1&{\rm line~}j{\rm ~crosses~}i{\rm ~from~left}\\
0&{\rm line~}j{\rm ~and~}i{\rm ~do~not~cross} \\
\end{array}
\right.
\ee 
Otherwise, $I_{ij}=0$. Note that $I_{ij}=-I_{ji}$.
Similarly, one can define the intersection matrix $J_{mv}$ of internal
and external lines. 

We shall now consider an arbitrary noncommutative graph $G$ in 
scalar NQFT \eq{nqft} and compute the corresponding contribution $I_G$
as given by noncommutative Feynman rules. We assume that $G$ has no tadpoles.
$\l(G)$ and $\V(G)$ denote the set of lines and vertices of the graph $G$
respectively. $I$ and $V$ denote the number of lines and vertices of $G$
respectively. Define the incidence matrix $\{\e_{vl}\}$, with indices
running over vertices and internal lines respectively, as
\be
\e_{vl}=\left\{ 
\begin{array}{cl}
1&{\rm if~the~vertex~}v{\rm ~is~the~starting~point~of~the~line~}l \\
-1&{\rm if~the~vertex~}v{\rm ~is~the~endpoint~of~the~line~}l  \\
0&{\rm if~}l{\rm ~is~not~incident~on~}v
\end{array}
\right.
\ee
Let us denote
by $P_v$ the total external momentum flowing into the vertex $v$. 
With these conventions
$I_G$ reads
\bea
I_G(P)&=&\int \prod_{l=1}^I d^d k_l \LB {1\ov k_l^2+m_l^2}\RB \prod_{v=1}^V 
[(2\pi)^d \delta^{(d)} (P_v -\sum_l \e_{vl} k_l)] \non
&&\ex \left[ i(\sum_{m,n} I_{mn} 
\theta_{\mu\nu}k_m^{\mu}k_n^{\nu}  + \sum_{m,v}  J_{mv} \theta_{\mu\nu}k_m^{\mu} P_v^{\nu})\right]
\la{IG}
\eea

\subsection{Parametric integral representation and topological formula}
The parametric integral representation of \eq{IG} is derived in appendix A
and it reads
\bea
I_G(P)&=&2^d (\sqrt{\pi})^{(I+V+1)d} \delta^{(d)}(\sum_v P_v)
\int_0^{\infty} \prod_{l=1}^I d\a_l  {{\rm e}^{-\sum_l \a_l m_l^2}\ov \sqrt{{\rm
det}\A {\rm det}\B}} \non
&& \ex\left\{ {1\ov 4} [\e \A^{-1} (J\eta)+2 i P]^{\mu}_v
(\B^{-1})^{\mu\nu}_{v\tilde v} [\e \A^{-1} (J\eta)+2 i P]^{\nu}_{\tilde v}
\right. \non
&&~~~~~
\left. -{1\ov 4} (J\eta)^{\mu}_m (\A^{-1})^{\mu\nu}_{mn} (J\eta)^{\nu}_n \right\} 
\la{alpha}
\eea
where
$$
\A_{mn}^{\mu\nu}\equiv \a_m \delta_{mn} \delta_{\mu\nu} - i I_{mn}
\theta_{\mu\nu},~~~~\B_{v\tilde v}^{\mu\nu}=\e_{vm}(\A^{-1})^{\mu\nu}_{mn}
\e_{\tilde v n},~~v,\tilde v=1,\ldots,V-1
$$
\be
(J\eta)^{\mu}_m\equiv\sum_{v=1}^V J_{mv}\eta^{\mu}_v,\, ~~~~~\eta^{\mu}_v\equiv
\theta_{\mu\nu}P_v^{\nu},\, ~~~~~~(y\e)^{\mu}_m\equiv \sum_{v=1}^V y_v^{\mu}\e_{vm}
\la{notation}
\ee

Without loss of generality one may assume that the matrix $\theta$ is
in the Jordan form
\begin{equation}\label{stc}
 \theta=\left(
 \begin{array}{ccccc}
 0 & -\theta_1 & & & \\
 \theta_1 &0& & & \\
  & &\ddots & & \\
 &&&0 & -\theta_{\frac{d}{2}}  \\
 &&&\theta_{\frac{d}{2}} &0 \\
 \end{array}
 \right).
 \end{equation}
Let $r$ be the rank of the intersection matrix $I_{ij}$. 
The structure of the pre-exponential factor in \eq{alpha} is then 
\be
\sqrt{\det\A \det\B}= \prod_{i=1}^{d/2} \LB \sum_{n=0}^{\g} 
\theta_i^{2n} P_{2n}(\a)\RB
\la{alpha1}
\ee
where $\g=r/2$ is defined in terms of the cycle number ${\tilde c}(G)$\footnote{For 
the definition of ${\tilde c}(G)$ see section 3. 
 $\g$ is simply the genus of the graph $G$ at vanishing external momenta.} as 
\be
\g=
{\tilde c}(G)/2
\la{gtilde}
\ee
and
$P_{2n}(\a)$ is a sum of monomials of degree $L-2n$: 
\be
P_{2n}=\sum_{\{ i_1,i_2,\ldots , i_{L-2n}\} }  \a_{i_1}\a_{i_2}\cdots 
\a_{i_{L-2n}} 
\la{alpha2}
\ee
Note that the coefficient in front of each monomial is one. 

\FIGURE[ht]{
\includegraphics[width=9truecm]{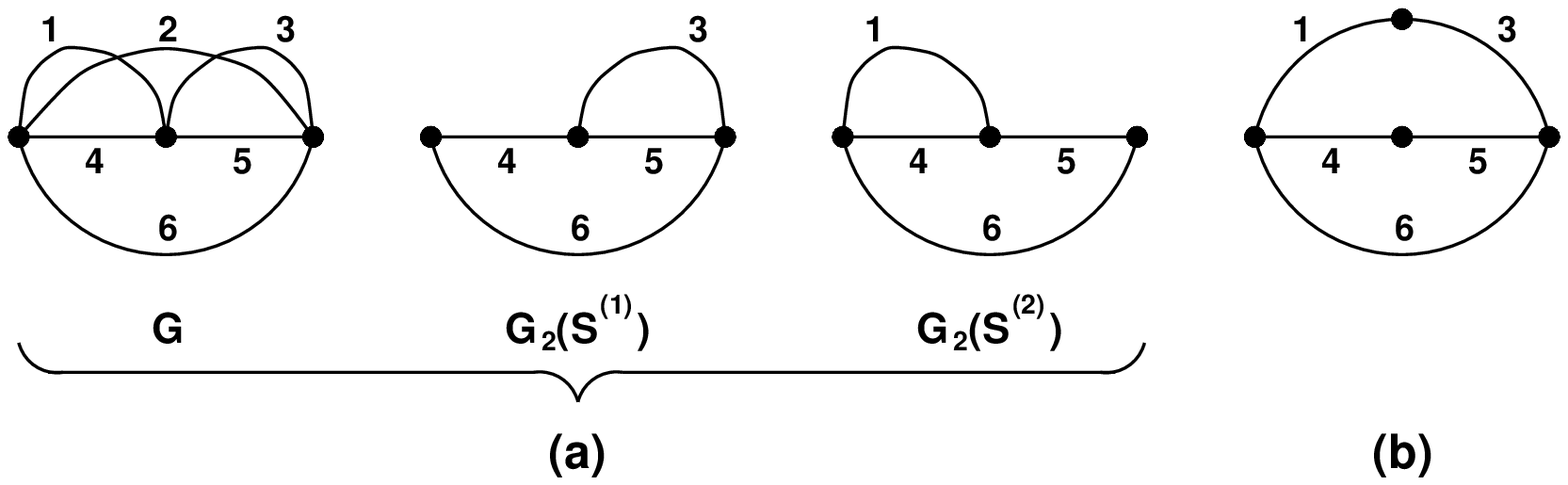}
\caption{Illustration for examples 3 and 5.}
}


We  now give a topological formula for $P_{2n}$. 
Let $\S_{2n}=\{ i_1,\ldots,i_{2n}\}\subset \R(G)$ be a set consisting of
$2n$ linearly independent lines from 
 $\R(G)$, i.e. the intersection matrix $I_{ij}$ 
restricted to
 the lines $i_1,\ldots,i_{2n}$ 
 is nondegenerate. 
For this set  $\S_{2n}$, 
one can define the graph $G_{2n}(\S)$ obtained from the graph
$G$ by deleting the lines of $\S_{2n}$.
Thus $\l(G_{2n}(\S))=\l(G)\setminus \S_{2n}$ and $\V(G_{2n}(\S))=\V(G)$. 
For a given graph $G$ and rosette $\R(G)$, there can be several 
different $\S_{2n}$: $\S^{(1)}_{2n},\ldots
\S^{(m)}_{2n}$. For each $\S^{(k)}_{2n}$ 
one has a graph $G_{2n}(\S^{(k)})$. 
For a given graph $\gamma$ define
the so-called chord-set product sum 
\be
C(\alpha,\gamma)=\sum_{\ell \in \T^*(\gamma)}\prod_{l\in \ell}\a_l
\la{chord}
\ee
where $\T^*(\gamma)$ is the set of all chords of the graph $\gamma$\cite{itzyk}. 

\noindent
{\it Example 3.} For the graph G in figure 6(a), 
$\S_2^{(1)}=\{ 1,2\}$, $\S_2^{(2)}=\{ 2,3\}$, and
$$C(\a,G_2(\S^{(1)}))=(\a_3+\a_5)(\a_4+\a_6)+\a_3\a_5$$
$$C(\a,G_2(\S^{(2)}))=(\a_1+\a_4)(\a_5+\a_6)+\a_1\a_4$$
Let us define the addition $\oplus$ on the space of  homogeneous 
polynomials of a given degree  
in $\alpha$'s  and with unit coefficients. Let $P_{2n}(\a)$ and $Q_{2n}(\a)$ be
two such polynomials. Then we define
$$P_{2n}\oplus Q_{2n}\equiv P_{2n}+Q_{2n} ({\rm mod~}2)$$

\noindent
{\it Example 4.}
$$(\a_1 \a_2+\a_2 \a_3+\a_4 \a_5)\oplus (\a_1\a_2+\a_4\a_5)\oplus (\a_4 \a_5+
\a_1\a_6)=\a_2 \a_3+\a_4\a_5+\a_1\a_6$$

With these conventions and definitions, the following theorem holds

\noindent
{\bf Theorem 1 (Topological formula).}   
\be
P_{2n}(\a,G)=\bigoplus_{k=1}^{m}  C(\alpha,G_{2n}(\S^{(k)}))  
\la{theorone}
\ee

The proof of this theorem is somewhat  technical and will not 
be given here. 
Note that for $n=0$, \eq{theorone} reads as $P_0(\a,G)=C(\a,G)$. 

\noindent
{\it Example 5.}
For the graph G in figure 6(a), we have 
\be
P_2=(\a_1+\a_3)(\a_4+\a_5)+\a_6 (\a_1+\a_3+\a_4+\a_5)
\la{ptwo}
\ee
Note that $P_2$ in \eq{ptwo} is equal to the chord-set product sum \eq{chord}
for the graph in figure 6(b).

\section{Convergence theorem}
This section is organized as follows. In subsection 3.1 we 
prove the convergence of Feynman integrals for some classes of 
graphs in the 
massive scalar NQFT. 
In subsection 3.2
we  formulate a general convergence theorem for noncommutative 
Feynman graphs and demonstrate it on the graphs
 discussed in subsection 3.1.

\subsection{Examples and propositions}

\FIGURE[ht]{
\includegraphics[width=9truecm]{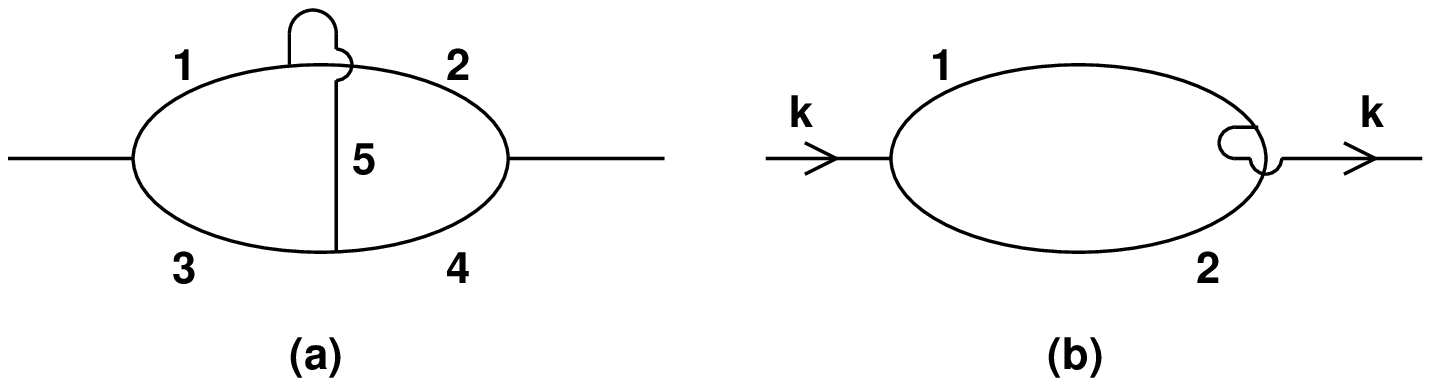}
\caption{}
}


In \eq{alpha} the UV divergences
show up as  poles at $\alpha=0$ of the integrand. The integral \eq{alpha}
is convergent at the upper limit of the integration because there are
no IR divergences in the massive theory.
Let us consider the diagram in figure 7(a). In the commutative limit it
is quadratically divergent in six dimensions. But as a noncommutative 
graph, it has a crossing of internal lines and it is
a genus $g=1$ graph. In $d$ dimensions, the prefactor of the 
exponent in \eq{alpha} for this
graph reads
\be
\sqrt{\det\A\det\B} =\prod_{i=1}^{d/2}[(\a_1+\a_3)(\a_2+\a_4)+\a_5(\a_1+\a_2+\a_3+\a_4)+
\theta_i^2]
\la{three}
\ee
Due to the non-zero $\theta$'s, \eq{three} has no zeros
in the range of integration of \eq{alpha}. Thus the graph in figure 7(a) is
convergent. It is easy to see from the $\alpha$-representation \eq{alpha}
that at  large external momentum $k$ it behaves as $\sim 1/k^{10}=
k^{\omega - 2g d}$ in any dimension $d$. Note that the graph in figure 5(a) is
an example of the graphs for which the number of lines in $\R(G)$ {\it
equals}  $2g$. The following proposition is true for such graphs.

\noindent
{\bf Proposition 1}
{\it If  the intersection matrix $I_{ij}(G)$ restricted to the lines of the 
rosette $\R(G)$
is nondegenerate and  $I_{G-\R(G)}$ for the planar
graph $G-\R(G)$ converges, then $I_G$ 
converges.}

\noindent
{\bf Proof.} Let $r=2g$ be the rank of the matrix $I_{ij}$.  Since $I_{ij}$ is
nondegenerate when restricted to the rosette, $m=1$ in  \eq{theorone}.
This
implies that 
$P_{2g}(G)=P_0(G-\R(G))$. Since
$$
{1\ov  \prod_{i=1}^{d/2} \LB \sum_{n=0}^{g} 
\theta_i^{2n} P_{2n}(\a)\RB} \le 
{1\ov \prod_{i=1}^{d/2} \LB  \theta_i^{2g} P_{2g}(\a)\RB }
$$
we conclude that $I_G$ converges if $I_{G-\R(G)}$ converges.~~~~{\bf q.e.d.}

Now consider the graph in figure 7(b). It has an intersection of the internal line with the
external line.  \eq{alpha} for this graph yields
\be
\int_0^{\infty}d\a_1 d\a_2 {1\ov (\a_1+\a_2)^{d/2}}{\rm exp}\LB
-(\a_1+\a_2)m^2-{\a_1\a_2 \ov \a_1+\a_2}k^2-{|\theta k|^2\ov 4(\a_1+\a_2)}\RB
\la{hook}
\ee
Note that in \eq{hook},  there is a term proportional to $1/\alpha$ in the
exponent.  
This is a general property of  graphs with   external lines
crossing  internal lines.  The $1/\alpha$ terms in the exponent will 
suppress  divergences coming from the pre-exponential factor. 
The following proposition is true for this type of graphs.

\noindent
{\bf Proposition 2}
{\it  If the non-planarity of a graph G is  solely due to the intersection of
an external line with one internal line as in figure 8., i.e. $J_{mv}=
\delta_{m m_0}\delta_{v v_0}$ and $I_{ij}\equiv 0$, and for any
1PI subgraph $\gamma \subset G$ {\it not containing line $m_0$} 
one has $\omega(\gamma)<0$, then 
$I_G$ converges.}

\FIGURE[ht]{
\includegraphics[width=5truecm]{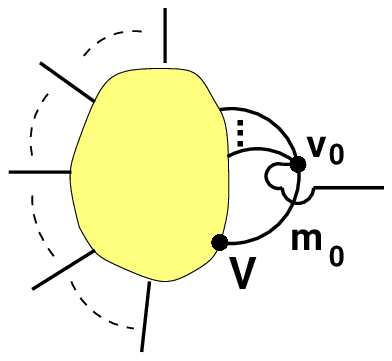}
\caption{}
}


\noindent
{\bf Proof.} At $\alpha_1=\cdots=\a_I=t \sim 0$, 
various terms in the exponent of \eq{alpha}   
scale as follows. The $J^0$ term scales like $O(t)$. The $J^1$ term gives
rise to an oscillatory contribution and thus cannot suppress the divergence
at $\alpha\sim 0$. The $J^2$ term scales like $O(1/t)$. We thus 
consider only $J^2$ term.

\noindent
Since $I_{ij}=0$ we have
$(\A^{-1})_{mn}^{\mu\nu}={1\ov \a_n} \delta_{mn}\delta_{\mu\nu}$. Let
us choose $V$ in \eq{notation} as in figure 8.  Simple
algebra gives
$$
(J\eta)^{\mu}_m (\A^{-1})^{\mu\nu}_{mn} (J\eta)^{\nu}_n={|\eta_{v_0}|^2\ov
\a_{m_0}}
$$
\be
[\e \A^{-1} (J\eta)]^{\mu}_v
(\B^{-1})^{\mu\nu}_{v\tilde v} [\e \A^{-1} (J\eta)]^{\nu}_{\tilde v}=
{|\eta_{v_0}|^2\ov \a_{m_0}^2}\B^{-1}_{v_0 v_0}
\la{jab}
\ee
Using 
\be
\B^{-1}={1\ov \det\B}\left. \det\B\right|_{v_0 {\rm ~deleted}}
\ee
and the fact that deleting $v_0$ is equivalent to shrinking the line $m_0$,
one finds 
\be
\B^{-1}_{v_0v_0}={\a_{m_0} P_0(G/m_0)\ov P_0(G)}
\la{jabt}
\ee
where $G/m_0$ denotes the graph obtained from $G$ by shrinking the line
$m_0$.

\noindent
Using \eq{jab}, \eq{jabt} and the relation 
\be 
P_0(G)=\a_{m_0} P_0(G-m_0)+P_0(G/m_0)
\ee
 it is easy to show that
the $J^2$ term in \eq{alpha}  
gives the following contribution
$$\ex \LB -{|\eta_{v_0}|^2 \ov 4}{P_0(G-m_0)\ov P_0(G)}\RB $$
Following ref.\cite{itzyk}, let us divide the integration domain in
\eq{alpha} into sectors
$$
0\le \a_{\pi_1}\le \a_{\pi_2}\le \cdots \le \a_{\pi_I}
$$
where $\pi$ is a permutation of $(1,2,\ldots,I)$. To each sector 
corresponds a family of nested subsets $\gamma_l$ of lines of G:
$$
\gamma_1\subset \gamma_2 \subset \cdots \subset \gamma_I = G
$$
where $\gamma_l$ contains the lines pertaining to $(\a_{\pi_1},\ldots,\a_{\pi_l})$. In the sector given by $\pi$, perform a change of variables
\be
\begin{array}{ccccccccc}
\a_{\pi_1}&=&\b_1^2&\b_2^2& \cdots &\b_s^2&\cdots&\b_{I-1}^2&\b_I^2 \\
\a_{\pi_2}&=&& \b_2^2 & \cdots &\b_s^2&\cdots&\b_{I-1}^2&\b_I^2 \\
~&\vdots& &&&&&& \\
\a_{\pi_s}&=&&&&\b_s^2&\cdots&\b_{I-1}^2&\b_I^2 \\
~&\vdots& &&&&&& \\
\a_{\pi_{I-1}}&=&&&&&&\b_{I-1}^2& \b_I^2 \\
\a_{\pi_I}&=&&&&&& &\b_I^2
\end{array}
\la{permute}
\ee
the jacobian of which is
$$
{D(\a_1,\ldots,a_I)\ov D(\b_1,\ldots,\b_I)}=2^I \b_1\b_2^3\cdots\b_I^{2I-1}
$$
In these $\beta$ variables the integration domain in \eq{alpha} reads
$$
0\le \b_I\le \infty~~~~~{\rm and}~~~~~0\le \b_l\le 1~~~~{\rm for}~~~~ 1\le l\le I-1
$$
Let $L_l$ be the number of independent loops in $\gamma_l$. It can
be shown that\cite{itzyk}
\be
P_0(G)=\b_1^{2L_1}\b_2^{2L_2}\cdots \b_I^{2L_I} [1+O(\b)]
\ee
We  derive a similar relation for $P_0(G-m_0)$ as follows.
Suppose that $\pi_s=m_0$. Then the graphs $\gamma_1,\ldots,\gamma_{s-1}$
do not contain line $m_0$. The graph $\gamma_s$ contains line $m_0$, but
it may happen that one end of $m_0$ is free i.e.  $m_0$ is attached
to  $\gamma_{s-1}$ with only one end. Let $\gamma_k,~k\ge s$ be the
first graph for which both ends  of  $m_0$ are not free.  By inspection
of \eq{permute} it is not difficult to see that
\be
P_0(G-m_0)=  \b_1^{2L_1}\b_2^{2L_2}\cdots \b_{k-1}^{2L_{k-1}} \b_k^{2(L_k-1)}
\cdots \b_I^{2(L_I-1)}
\ee
The leading term at $\alpha\sim 0$ in the integrand of \eq{alpha} is
then
\bea
&&
{\b_1^1 \b_2^3 \cdots \b_I^{2I-1}\ov \b_1^{dL_1}\b_2^{dL_2}\cdots \b_I^{dL_I}}
\ex\LB -{|\eta_{v_0}|^2\ov 4}{1\ov \b_k^2 \cdots \b_I^2}\RB \non
&&=
\LB \prod_{l=1}^{k-1} \b_l^{-\omega_l-1}\RB \LB \prod_{l=k}^I \b_l^ {-\omega_l-1}\RB \ex\LB -{|\eta_{v_0}|^2\ov 4}{1\ov \b_k^2 \cdots \b_I^2}\RB
\eea
where $\omega_l=dL_l-2 l$. Since $\omega_l<0$ for $l<k$, the integral
\eq{alpha} converges.~~~~{\bf q.e.d.}

Let us mention a peculiar feature of the diagrams with  external
lines crossing internal lines(figure 7(b)). 
At large external momenta they scale like 
$\ex (-{\rm const.~}k^2 \theta)$.\footnote{This UV behavior of diagrams
may have relation to the UV behavior of gauge non-invariant
correlators of large-N noncommutative SYM theory calculated in 
ref.\cite{russo} using supergravity/gauge theory correspondence.}
But once such a graph is put inside a bigger graph, it behaves as if it 
has dimension $\omega - d$.

\noindent
{\it Example 6.} The subgraph $\gamma$ formed by  lines $1$ and $4$ 
of  graph G in figure 6(a) has line $2$ as an external line 
crossing the internal line 1. A simple rescaling 
 $\alpha_1,\a_4 \rightarrow \rho\alpha_1,\rho\a_4$ in 
$$
{d\a_1 \cdots d\a_6 \ov (P_0+\theta^2 P_2)^{d/2}}$$
where $P_2$ is given by
 \eq{ptwo}, shows
that the subgraph $\gamma$ behaves as $k^{\omega(\gamma) -d}$ when its  external
momenta are large.  

\noindent
This circumstance of a graph behaving differently in 
different
contexts makes it difficult to implement  the approach of the asymptotic
algebra developed in refs.\cite{weinberg,kenedy} for the analysis
of the asymptotic behavior of usual Feynman diagrams to our case. 
Presumably, one may find an  algebra of asymptotic 
functions in  the NQFT case and use it for an inductive proof of 
the convergence theorem.

In the remaining  example  and proposition  of this subsection we will
need  
the following lemma. 

\noindent
{\bf Lemma }{\it Let us choose $n$ vertices of a graph $G$ and identify
$n-1$ of them. Let $j$ be the remaining vertex. Denote by $G_j$ the resulting
graph. Letting $j$ to run from 1 to $n$ one finds different $G_j$'s.
Then the following relation holds
 $$\bigoplus_{j=1}^n P_0(G_j)=0$$
Pictorially, it reads   
\be
\bigoplus_{j=1}^nP_0(\picl{4}{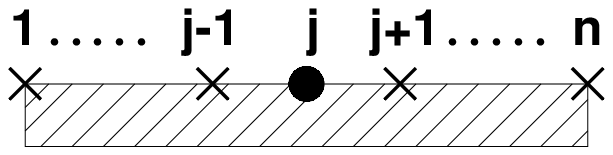})=0
\la{identity}
\ee
where the crosses $\times$   denote identified vertices.}

\noindent
The proof of this lemma is given in appendix B.  

The relations 
proved in the following example and proposition will be used in
subsection 3.2 for the analysis of the convergence properties of
the  graphs.

\noindent
{\it Example 7.} Consider the graph $G_1$: $\pic{4}{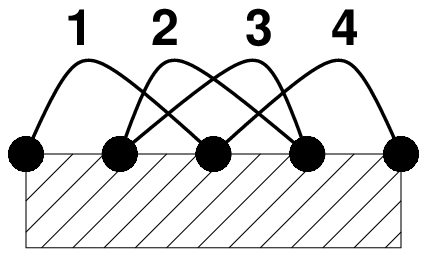}$. Let us prove that 
\be
P_2(\pic{4}{ex1.ps})=P_0(\pic{4}{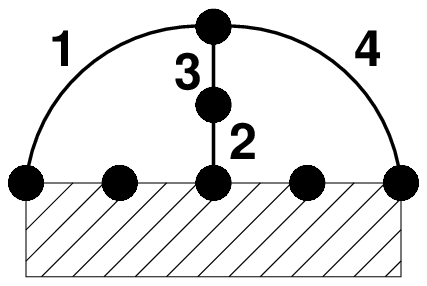})
\la{exam1}
\ee
where the hashed block denotes an arbitrary planar subgraph. 
$G_1$ is a $g=1$ graph and it is not of the type 
considered in Proposition 1. 
Applying Theorem 1 to the LHS of \eq{exam1} one finds
\bea
P_2(\pic{4}{ex1.ps})=P_0(\pic{4}{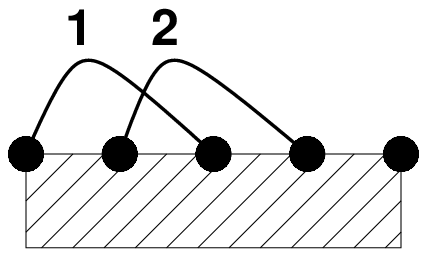})\oplus P_0(\pic{4}{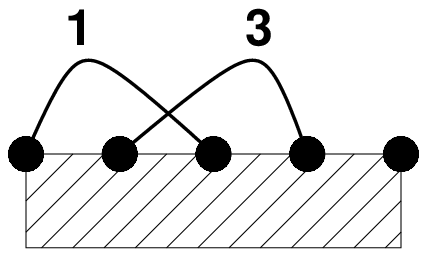})
&\oplus&  P_0(\pic{4}{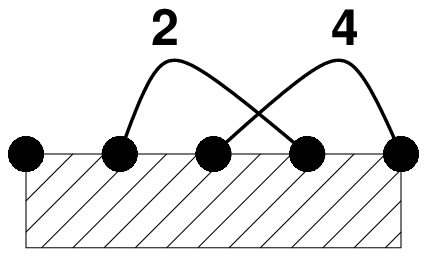})\oplus P_0(\pic{4}{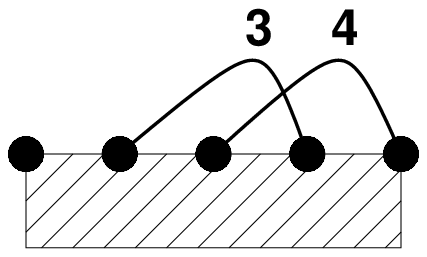})
\oplus\nonumber\\ 
&\oplus&  P_0(\pic{4}{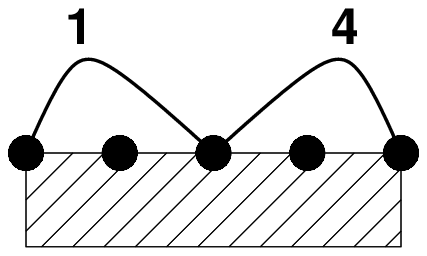})
\eea
Let $G/l$ and $G-l$ be the  graphs obtained from $G$ by shrinking and deleting
the line $l$ respectively. 
Using the general relation $P_0(G)=\a_l P_0(G-l)+P_0(G/l)$, we have
\bea
P_0(\pic{4}{exg1.ps})&=&\alpha_2 P_0(\pic{4}{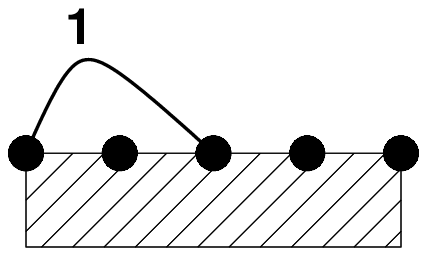})+
P_0(\pic{4}{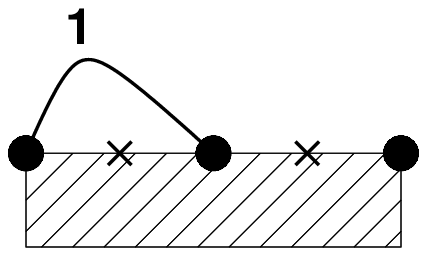}) \non
P_0(\pic{4}{exg2.ps})&=&\alpha_3 P_0(\pic{4}{exg12.ps})+
P_0(\pic{4}{excg12.ps})\non
P_0(\pic{4}{exg3.ps})&=&\alpha_2 P_0(\pic{4}{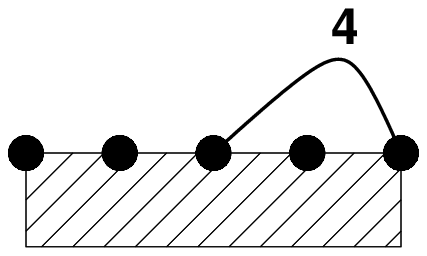})+
P_0(\pic{4}{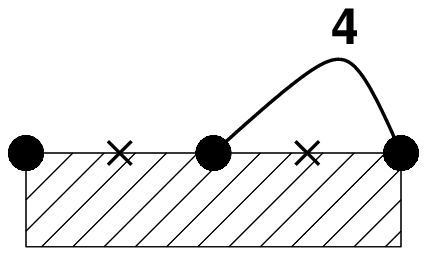}) \non
P_0(\pic{4}{exg4.ps})&=&\alpha_3 P_0(\pic{4}{exg34.ps})+
P_0(\pic{4}{excg34.ps})
\la{exam2}
\eea
Using \eq{exam2} and the definition of $\oplus$ one finds
\bea
P_2(\pic{4}{ex1.ps})&=&
(\alpha_2 + \alpha_3)\Big[P_0(\pic{4}{exg12.ps})\oplus 
P_0(\pic{4}{exg34.ps})\Big] \oplus  P_0(\pic{4}{exg5.ps}) \non
&=&(\alpha_2 + \alpha_3)\Big[(\alpha_1+\alpha_4)P_0(\pic{4}{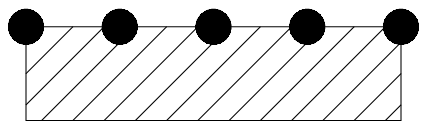})
+\big(P_0(\pic{4}{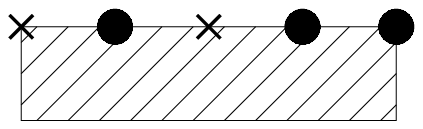})\oplus P_0(\pic{4}{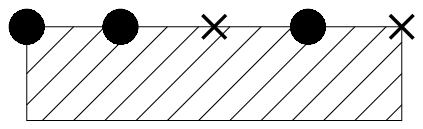})\big)\Big]\non
&&\oplus  P_0(\pic{4}{exg5.ps})\non
&=&(\alpha_2 + \alpha_3)\Big[(\alpha_1+\alpha_4)P_0(\pic{4}{i2ex1.ps})
+P_0(\pic{4}{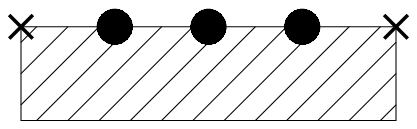})\Big]\non
&&+  P_0(\pic{4}{exg5.ps}) \non
&=&(\alpha_2+\alpha_3)P_0(\pic{4}{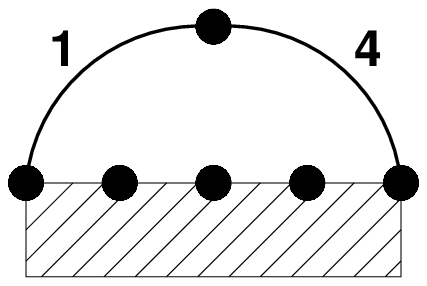}) +P_0(\pic{4}{exg5.ps}) \non
&=&P_0(\pic{4}{iex1.ps})
\eea
where for the third equality  we used the Lemma.
Thus we  have proven  \eq{exam1}. 
  \eq{exam1} will be used in section 3.2
for the  analysis of the convergence property of $G_1$.  
The reader may try to prove  similar relations for various 
graphs involving several crossing lines. One can even prove some 
quite general relations as in
the following proposition.

\noindent
{\bf Proposition 3}
\be
P_{2g}(\,\picel{6}{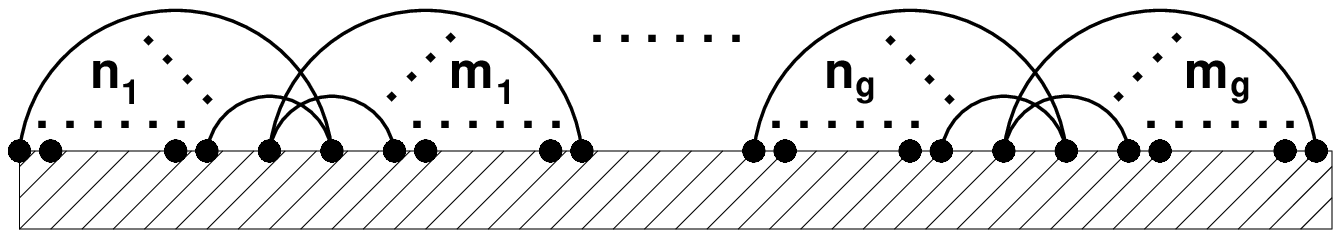}\,)=P_0(\,\picel{6}{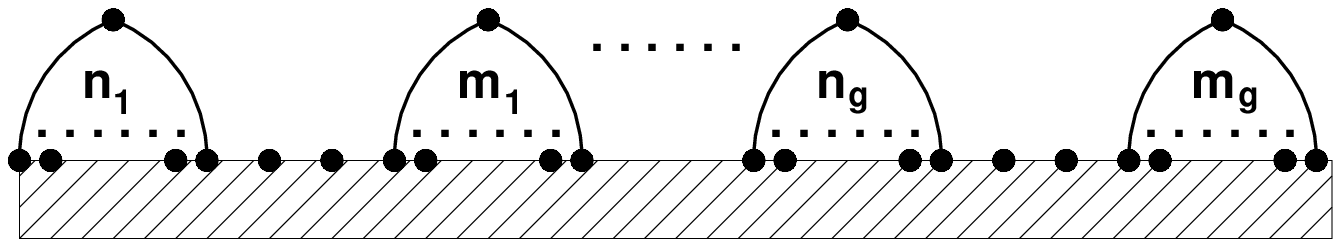}\,)
\la{proptre}
\ee
\noindent
This proposition can be proven by induction using the Lemma  along the 
lines of 
the proof given in
the example 7. 

A remarkable feature of  these relations is that
they relate
$P_{2g}$ of a genus $g$ graph  to $P_0$ of a genus zero graph. 
This suggests that there should exist a {\it natural} map 
$${\cal F}_{g_1\rightarrow g_2}:{\cal G}_{g_1}\rightarrow {\cal G}_{g_2},~~
g_1>g_2$$
between sets ${\cal G}_{g_1}$, ${\cal G}_{g_2}$ of graphs   
 of  genera $g_1$, $g_2$, such that the  relation 
$$
P_{2g_1}(G_{g_1})=P_{2g_2}({\cal F}_{g_1\rightarrow g_2}(G_{g_1}))
$$
where $G_{g_1}\in {\cal G}_{g_1}$, holds.

\subsection{Convergence theorem and analysis of various graphs}

In this subsection we  formulate a general convergence theorem for
the noncommutative graphs and illustrate it on the graphs considered
in section 3.1.  Let us  give several definitions required for
the formulation of the theorem.  
Let $G$ be a genus $g$ graph with a set of external lines $\E(G)$. 
Such a graph can be drawn on a genus $g$  2-surface, $\Sigma_g$, 
with a boundary, $\partial\Sigma_g$, to which the external lines are 
attached.

Let us  define a
cycle number $c(\gamma)$ for the subgraph $\gamma\subset G$.  
The first homology group of  $\Sigma_g$ for the graph $G$ has the basis 
${\cal C}=\{ a_1,b_1,\ldots,a_g,b_g\}$ (see figure 1). One may go to a 
different basis by forming combinations of the elements of ${\cal C}$.
$c(\gamma)$ is defined as the number of inequivalent non-trivial cycles
of $\Sigma_g$ spanned by the closed paths in $\gamma$. The 
following example illustrates the definition of $c$.

\noindent
{\it Example 8.}  Let us denote the hashed planar part of  graph $G_1$ in 
example 7 by $\gamma$. If we  draw  $G_1$ on a $g=1$ surface, we 
see that  $c(\gamma \cup \{ i\})=1, i=1,\ldots,4$.

Let us consider now a graph $G$ at the vanishing $p=0$  external momenta.
It can be drawn on  a genus $g'$ 2-surface 
$\Sigma_{g'}^{p=0}$. In general $\Sigma_{g'}^{p=0}$ is different from
$\Sigma_g$.
Let us  define the cycle number ${\tilde c}(\gamma)$  of the subgraph 
$\gamma \subset G$ to be the number of inequivalent non-trivial
cycles of $\Sigma_{g'}^{p=0}$ spanned by the closed paths in $\gamma$. 
In general $c(\gamma) \ne {\tilde c}(\gamma)$.

\FIGURE[ht]{
\includegraphics[width=12truecm]{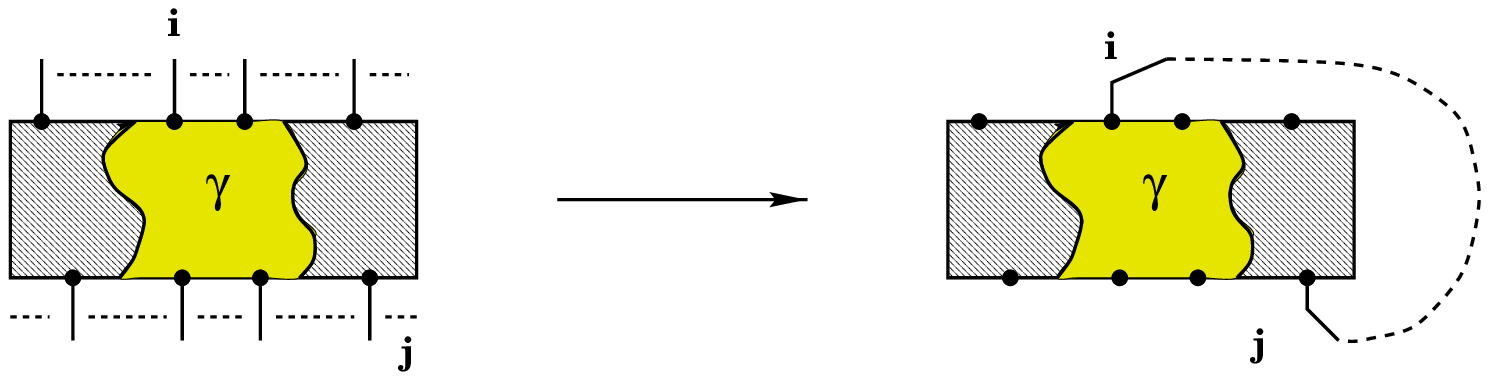}
\caption{Definition of index $j$.}
}

Consider two external lines  $i,j  \in  \E(G)$.
As in figure 9, set the rest of the external momenta graph $G$ to 
zero and connect 
the lines $i$ and $j$. Let $c_{ij}(\gamma)$ be the cycle number of 
an arbitrary subgraph $\gamma \subset G$  with respect to 
the 2-surface of the resulting graph. There are only two possibilities:
 $$c_{ij}(\gamma) > {\tilde c}(\gamma)~~~~~{\rm or}~~~~~
 c_{ij}(\gamma)={\tilde c}(\gamma)$$
We define the index $j$ of an arbitrary (connected or disconnected ) subgraph
 $\gamma \subset G$ as 
follows:  if there {\it exists} a pair of external lines $i, j$ such that
$c_{ij}(\gamma) > {\tilde c}(\gamma)$, then $j(\gamma)=1$. 
Otherwise, $j(\gamma)=0$. The following examples illustrate
this definition.

\noindent
{\it Example 9(a).} Consider the following diagram:
\FIGURE[ht]{
\includegraphics[width=4truecm]{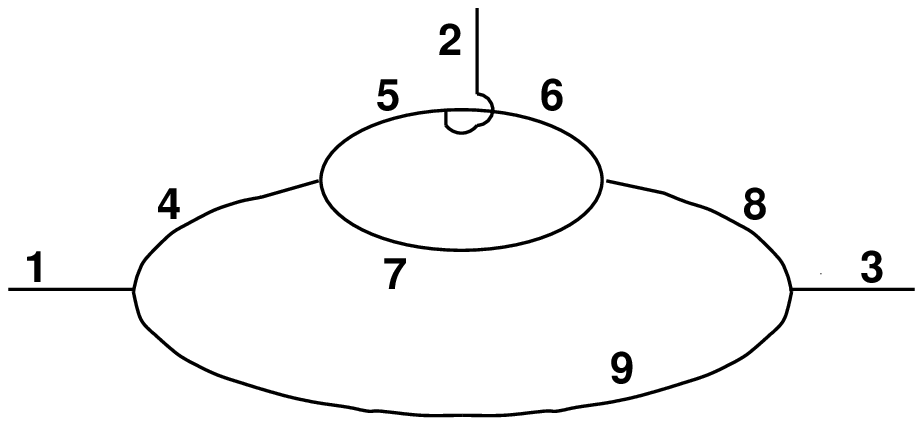}
\caption{}
}

This graph has genus $g=1$ and $c(G)=1$. By setting the external momenta 
to zero we get $g'=0$ and ${\tilde c}(G)=0$. By joining line $2$ with 
line $1$ we get $c_{21}(G)=1$ (we also have $c_{23}(G)=1$).  Thus, according 
to our definition, $j(G)=1$. Let us analyze the subgraphs of $G$:

\noindent
$\bullet\,\,\,\,\gamma_1=\{5, 6, 7\}$: this subgraph has
$c_{21}(\gamma_1)=c_{23}(\gamma_1)=1$ and ${\tilde c}(\gamma_1)=0$
which implies $j(\gamma_1)=1$

\noindent
$\bullet\,\,\,\,\gamma_2=\{4, 7, 8, 9\}$: this subgraph 
does not wrap any non-trivial cycle; $c(\gamma_2)=0={\tilde c}(\gamma_2)$
and thus $j(\gamma_2)=0$

\noindent
$\bullet\,\,\,\,\gamma_3=\{4, 5, 6, 8, 9\}$: this subgraph has properties
similar to $\gamma_1$: $c_{21}(\gamma_3)=c_{23}(\gamma_3)=1$ and 
${\tilde c}(\gamma_3)=0$ and therefore $j(\gamma_3)=1$
\smallskip

\noindent
{\it Example 9(b).} Consider the graph in figure 10. It has genus $g=2$ and 
$c(G)=3$. This is also the $g$ and $c$ for the graph obtained by joining
the external lines. It is easy to see that by setting the external momenta 
to zero the genus becomes $g'=1$ and ${\tilde c}(G)=2$. We therefore conclude 
that $j(G)=1$.

\FIGURE[ht]{
\includegraphics[width=3truecm]{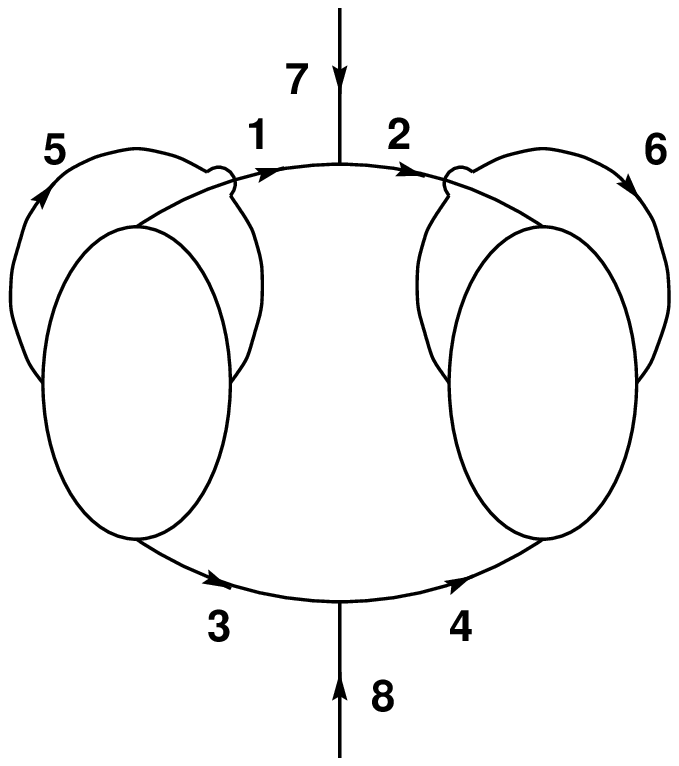}
\hspace{2truecm}
\includegraphics[width=6truecm]{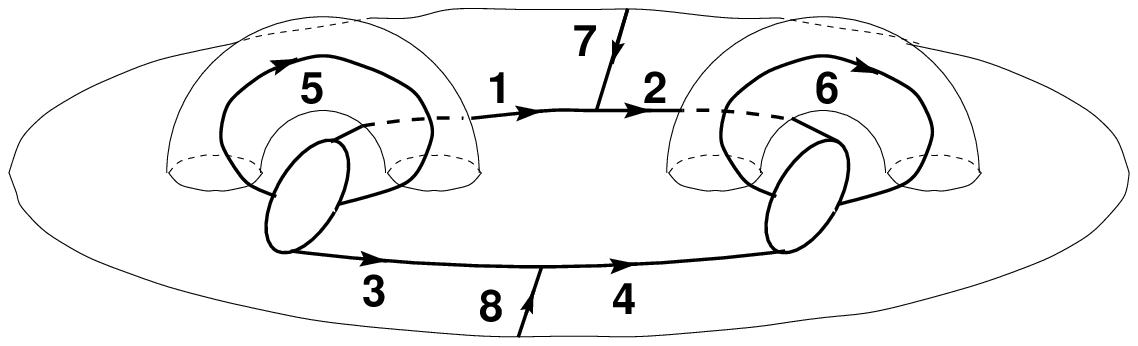}
{\vspace{2truemm}}
\centerline{~~~~~~~{\bf (a)}~~~~~~~~~~~~~~~~~~~~~~~~~~~~~~~~~~~~~~~~~~~~~~~~~~{\bf (b)}~~~~~~~~~~~~~~~~}
{\vspace{-8truemm}}
\caption{}
}
Let us explain the difference between  homologically trivial and
nontrivial cycles from the point of view of the momentum flow
on the surface $\Sigma_g$ of a graph. 
Figure 1(b) illustrates the
flow of momentum on a genus two surface with a boundary.
There are  {\it topologically trivial} flows like $p_0$ and 
{\it topologically nontrivial} flows like $p_A$, $p_C$, $p_{a_2}$
and $p_{b_2}$. 
Since 
the total external momentum flowing into the surface $\Sigma_2$ through 
$\partial\Sigma_2$ is zero,  the net momentum flowing across $A$ and
$C$ is zero (the momenta $p_A$, $p_C$ along $A$ and $C$ are in general
nonzero). 
The  phase factor associated with  a graph arises from the {\it
linking} of topologically nontrivial flows. In figure 1(b), $p_{a_2}$
and $p_{b_2}$  contribute  a phase factor 
$\ex (i\theta_{\mu\nu}p_{a_2}^{\mu}p_{b_2}^{\nu})$.
Cycles $A$ and $C$ do not contribute to the phase factor because
the net momentum flowing across each of these cycles is zero.
Since the cycles $a_2,b_2$ are homologically nontrivial and the
cycles $A$, $C$ and $0$ are homologically trivial, we conclude
that 
\noindent
{\it only the momentum flow along the homologically 
nontrivial cycles contribute to the
phase factor.}

\noindent
{\it Example 10(a).} The total momentum flowing along the cycle $a$ of the
graph in figure 2(b) is $p_a=q_8+q_{10}$. The total momentum
flowing along the cycle $b$ is $p_b=q_8+q_{10}-q_4$. Thus
the phase factor is 
$$\ex (i\theta_{\mu\nu}p_{a}^{\mu}p_{b}^{\nu})=
\ex (i\theta_{\mu\nu}q_{4}^{\mu}(q_8+q_{10})^{\nu})
$$

\noindent
{\it Example 10(b).} In figure 11(b) $p_{b_1}=q_5$, $p_{b_2}=q_6$,
$p_{a_1}=-q_1$ and $p_{a_2}=q_2$. Thus the phase factor is
$$
\ex (i\theta_{\mu\nu} (q_5^{\mu} q_1^{\nu} +q_2^{\mu} q_6^{\nu}))
$$

One might  object that the  statement made above regarding the 
homologically nontrivial cycles is 
not always true by giving the following  counter-example.
In figure 12(a)  a graph $G$ is drawn on a genus $g$ surface with a boundary.
The subgraph $\gamma$  wraps the cycle $a_g$  and it is connected
to the rest of the diagram only through the handle $g$. Due to
the momentum conservation, the net momentum flowing along the
cycle $b_g$ is zero. Thus there is no phase factor  associated with
$\gamma$. The subgraph $\gamma$ seems to be  homologically nontrivial,
but there is no phase factor associated with it. 
The point is that one
can ``slide''  $\gamma$ through the handle $g$ and redraw the 
2-surface as in figure 12(b). 
\FIGURE[ht]{
\includegraphics[width=7truecm]{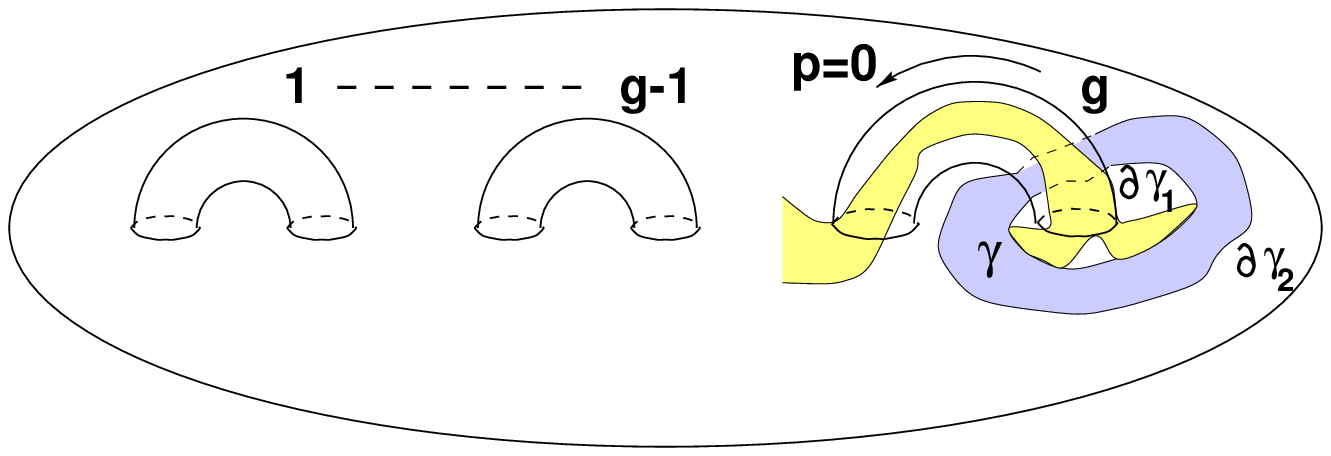}
\hspace{.5truecm}
\includegraphics[width=7truecm]{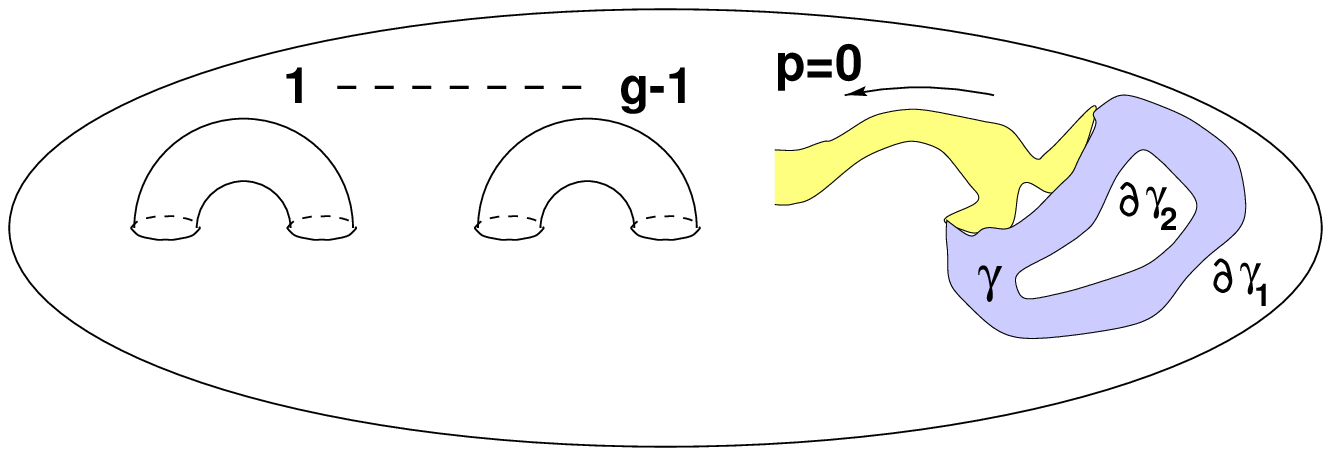}
{\vspace{2truemm}}
\centerline{~~~~~~~~~~~~{\bf (a)}~~~~~~~~~~~~~~~~~~~~~~~~~~~~~~~~~~~~~~~~~~~~~~~~~~~~~~~~{\bf (b)}~~~~~~~~~~~~~}
{\vspace{-8truemm}}
\caption{Two equivalent graphs.}
}
The resulting surface has genus $g-1$.
Considered as noncommutative Feynman graphs, the graphs in 
figure 12(a) and figure 12(b)
are the same graph, i.e. it appears only once in the perturbative 
expansion.

\noindent
{\bf Theorem 2.(Convergence Theorem)}{\it In a massive NQFT in $d$ dimensions, 
a  1PI graph  $G$ is convergent if
and only if for any subgraph
\footnote{
In particular $\gamma$ can be a disjoint union of
1PI subgraphs of $G$. It is assumed that the external momenta of the graph
$G$ are generic. The integral may diverge for exceptional external momenta
which form a set of measure zero in the space of external momenta.} 
 $\gamma\subseteq G$ at least one of
the following conditions is satisfied: 
\begin{enumerate}
\item $\omega(\gamma)-c(\gamma)d< 0$ 
\item $j(\gamma)=1$.
\end{enumerate}
}

\noindent
An inductive proof of this theorem will be given in ref.\cite{chep}.

It is not difficult to see from \eq{alpha},\eq{alpha1},
\eq{alpha2} and the relation
$$
P_{2n}(G)=P_{2n}(\gamma)P_0(G/\gamma)+X
$$
($X$ is a sum of  terms whose degree with respect to
$\a_l~(l\in \gamma)$ is at least $L(\gamma)-2n+1$), that the conditions (1) and (2) are necessary. The non-trivial part of the convergence theorem is
the {\it sufficiency} of the conditions (1)-(2).

We  now demonstrate that  Theorem 2 holds for the graphs 
we considered in previous sections. 

\noindent
{\it Analysis of figure 6}

The condition (1) is satisfied for any subgraph
of $G$ if $d<6$. Using  the relation $1/(P_0+\theta^2 P_2)\le 1/\theta^2P_2$
and   the fact that $P_2(G)$ equals $P_0$ of the graph shown in
figure 6(b), we see that $I_G$ indeed converges if $d<6$.

\noindent
{\it Analysis of  Proposition 1}

Proposition 1 states that
$I_G$ converges if $I_{G-\R(G)}$ converges. Let us see how this follows
from  Theorem 2.
The subgraph $G-\R(G)$ is planar and according to the condition (1) of
Theorem 2  it should satisfy $\omega(G-\R(G))<0$.
Since
$$
\omega(G)=\omega(G-\R(G))+(d-2)I(\R)= \omega(G-\R(G))+2g(d-2)
$$   
the condition $\omega(G)-2g d<0$ is satisfied. One can similarly show
that for any non-planar subgraph of $G$ the condition (1) is satisfied if 
 $\omega<0$ for all planar subgraphs of $G$.

\noindent
{\it Analysis of Proposition 2}

Choose the decomposition of the external lines $\E(G)=\{P_{v_0}\}\cup
 \{ {\rm rest}\}$. Thus $j(G)=1$. Any subgraph of $G$ not containing
the external line $P_{v_0}$ has $j=0$ and thus should satisfy $\omega<0$.
Any subgraph of $G$ which contains $P_{v_0}$ and at least one other
external line of $G$ has $j=1$. Such a subgraph may have $\omega\ge 0$,
but it satisfies the condition (2) of the convergence theorem.

\noindent
{\it Analysis of example 7}

Let us denote the hashed block of $G_1$ as $\gamma$. 
From the relation $\omega(G_1)=\omega(\gamma)+4(d-2)$, we see
that the condition (1) for $G_1$: $\omega(G_1)-2d<0$ is satisfied
if $\omega(\gamma)<8-2d$. Let us see how the same
conclusion follows from \eq{exam1}. \eq{exam1} states that 
$P_2(G_1)=P_0(G_0)$. Thus $I_{G_1}$ converges if $I_{G_0}$ is convergent.
One of the conditions for the covergence of $I_{G_0}$ is 
$\omega(G_0)<0$, or equivalently, $\omega(\gamma)+2d-8<0$.
Let us consider the subgraph $\gamma \cup \{ 1,2,3\}$ next.
For this subgraph $c(\gamma \cup \{ 1,2,3\} )=2$.
Condition (1) of Theorem 2 says that the subgraph should satisfy
$\omega(\gamma)+3(d-2)-2d<0$ or, equivalently, $\omega(\gamma)<6-d$.
The same restriction follows from \eq{exam1}, since
the degree of divergence of the subgraph 
$\gamma \cup \{ 1,2,3\} \subset G_0$ is  $\omega=\omega(\gamma)+d-6$.   
For $d\ge 2$ we have $6-d\ge 8-2d$. Thus we are left with a single
condition $\omega(\gamma)<8-2d$. One may derive analogous relations
for the subgraphs of $\gamma$. 

\noindent
{\it Analysis of Proposition 3}

Let us denote the hashed block in \eq{proptre} by $\gamma$. 
The subgraph of $G_{g}$ formed by $\gamma$ and the lines $1,\ldots,n_1$
has $c=1$ and $\omega =\omega(\gamma)+n_1(d-2)$. The condition (1) of
the Theorem 2  reads as $\omega(\gamma)+n_1(d-2)-d<0$. Let us inspect
the graph $G_0$ on the RHS of \eq{proptre}. The subgraph of $G_0$ formed
by $\gamma$ and the lines $1,\ldots,n_1$ has 
$\omega=\omega(\gamma)+n_1(d-2)-d$. Thus we have the same convergence 
condition that
we found before.

\section{Subtraction of divergences and counterterms}
\subsection{Subtraction of divergences}
In Section 3 we argued that the convergence theorem  holds for the
noncommutative scalar theories with non-derivative couplings. 
In what
follows we assume that it holds also for theories with derivative
couplings.

In this section we propose a noncommutative analog of Bogoliubov-Parasiuk's
recursive subtraction formula and show that it leads to finite
integrals. Our discussion will be  parallel
to one in the commutative QFT case and we refer the readers 
not familiar with the
subject of BPHZ renormalization 
to  the
ref.\cite{itzyk} for a nice and elementary introduction.

For the reason given at the beginning of  Section 1.2, for a
particular NQFT in $d$ dimensions, we restrict
our discussion to the graphs  of   class $\Omega_d$. 
The class $\Omega_d$ consists of  graphs 
whose  {\it topologically nontrivial} subgraphs satisfy at least 
one of the conditions (1)-(2) of Theorem 2.  By definition, a subgraph 
$\gamma \subseteq G$ is topologically nontrivial (=nonplanar) if  on 
$\Sigma_g$ none of the closed paths in $\gamma$ can be
contracted to a point. Note that  a topologically nontrivial
graph is not necessarily homologically nontrivial.\footnote{See footnote 7.}
Topologically nontrivial, but homologically trivial subgraphs have
$c=0$ and so the condition (1) of Theorem 2 for such graphs reads as
$\omega<0$.
 
This means that if $G\in \Omega_d$, then only topologically trivial subgraphs 
of $G$ are
allowed to violate the conditions (1)-(2) of Theorem 2. Our subtraction
procedure renders the graphs from the class $\Omega_d$ finite. 
We will show that the recursion formula applied  to
the integrand $\I_G$ of a graph $G\in \Omega_d$ yields  an expression which
satisfies the  conditions of the convergence theorem.

Let $\Sigma_g$ be a particular genus $g$ surface on which the 
graph $G$ is drawn. There will be momenta flowing in the loops of
the diagram, but
as  pointed out in section 3.2, only the momentum
flow along the homologically nontrivial cycles contribute to the phase
factor.  
Let $p_{a_i}$, $p_{b_i}$ be the momenta flowing along the nontrivial 
cycles of $\Sigma_g$. 
Denoting by $\varphi_G$ the phase factor for  graph $G$, 
the  general form of the integrand $\I_G$ of  graph $G$ in 
momentum space reads
\be
\I_G(k,q,p_{a_i},p_{b_i})={\rm e}^{i\varphi_G(k,p_{a_i},p_{b_i})} 
\I_G^{\theta=0}(k,q,p_{a_i},p_{b_i})
\la{iofg}
\ee
where $\I_G^{\theta=0}$ is the integrand for the corresponding 
commutative QFT, and $k$ and $q$ denote the external and the rest of
independent loop momenta respectively. If  graph $G$ is planar,
then its phase factor $\varphi(k)$  depends only on the
 external momenta (see \eq{varphi}). 
Let us define  $\I_G$ for a planar graph $G$ to be
\be
\I_G=\I_G^{\theta=0}
\ee
and $\I_G$ for a nonplanar graph to be given by \eq{iofg}.

For a topologically trivial graph $G$, 
let us denote by  $R^{(0)}_G$ the renormalized  integrand 
that leads to a finite integral.
Let us denote by ${\bar R}^{(0)}_G$ the integrand with all subdivergences
except the overall divergence of $G$ subtracted.
Let $T_G$ be the  operator 
which acts on  ${\bar R}^{(0)}_G$ of a planar graph $G$ as 
follows. $T_G{\bar R}^{(0)}_G$ is
the Taylor expansion of ${\bar R}^{(0)}_G$ 
in the external momenta at the origin,
up to the order $\omega(G)$ included. Let $\Re(G)$ be the set of all
renormalization parts of the graph $G$, where by the renormalization part
we mean any planar 1PI subgraph $\gamma\subset G$ except for $G$ itself 
such that $\omega(\gamma)\ge 0$. When two subgraphs $\gamma_1$ and $\gamma_2$
have no common vertex nor line, we denote $\gamma_1\cap \gamma_2=\emptyset$.
With these conventions,
the recursive
subtraction formula for a  planar graph $G$ reads
$$
R^{(0)}_G=\left\{
\begin{array}{cl}
{\bar R}^{(0)}_G&{\rm ~if~}\omega(G)<0
\\
(1-T_G){\bar R}^{(0)}_G&{\rm ~if~}\omega(G)\ge 0 
\end{array}
\right.
$$
\be
{\bar R}^{(0)}_G=\I_G+\sum_{{\{ \gamma_1,\ldots,\gamma_s\}\atop 
\gamma_j \in \Re(G),~ 
\gamma_i\cap \gamma_j=\emptyset}}
\I_{G/{\{ \gamma_1,\ldots,\gamma_s\}}}\prod_{a=1}^s(-T_{\gamma_a}
{\bar R}^{(0)}_{\gamma_a})
\la{rzero}
\ee

Let us now consider a topologically nontrivial graph $G$.
Let  $\gamma_a \in \Re(G),~a=1,\ldots, s$, be a set of
 disjoint,
$\gamma_i\cap\gamma_j=\emptyset$,  renormalization parts. Since $\gamma_i$ 
are topologically trivial we have 
$$
\I_{\gamma_i}=\I_{\gamma_i}^{\theta=0}
$$
The integrand of the reduced graph $G/{\{ \gamma_1,\ldots,\gamma_s\}}$ reads
\be
\I_{G/{\{ \gamma_1,\ldots,\gamma_s\}}}=
{\rm e}^{i\varphi_G(k,p_{a_i},p_{b_i})} 
\I_{G/{\{ \gamma_1,\ldots,\gamma_s\}}}^{\theta=0}
\ee
\FIGURE[ht]{
\includegraphics[width=7truecm]{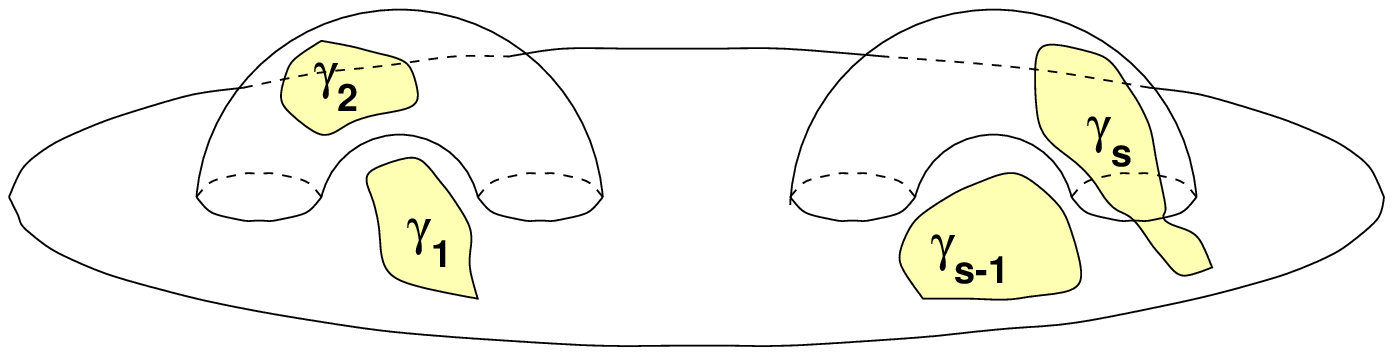}
\caption{Topologically trivial disjoint renormalization parts.}
}
The meaning of this  equation is the following. When
we shrink the renormalization parts $\gamma_1,\ldots,\gamma_s$, 
 the local structure of the graph $G$ changes, but the global structure
does not change because $\gamma_i$'s are topologically trivial and disjoint.
Thus the global flow of  momentum remains unchanged, implying
that the phase factor of the reduced graph $G/{\{ \gamma_1,\ldots,\gamma_s\}}$
is the same as that of $G$.

For a topologically nontrivial graph $G\in \Omega_d$, we define the
renormalized integrand $R^{(1)}_G$ that leads to a finite integral
 as follows (see figure 13).
\be
 R^{(1)}_G=\I_G+\sum_{{\{ \gamma_1,\ldots,\gamma_s\}\atop 
\gamma_j \in \Re(G),~ 
\gamma_i\cap \gamma_j=\emptyset}}
\I_{G/{\{ \gamma_1,\ldots,\gamma_s\}}}\prod_{a=1}^s(-T_{\gamma_a}
{\bar R}_{\gamma_a}^{(0)})
\la{subtract}
\ee
where ${\bar R}_{\gamma_a}^{(0)}$ is given by \eq{rzero}. Note 
that $R^{(1)}$ does not enter into the recursion, whereas $R^{(0)}$ does.
In other words \eq{rzero} is recursive, whereas \eq{subtract} is 
non-recursive.
 
For a general graph $G\in \Omega_d$ define
\be
R_G=\left\{
\begin{array}{cl}
R^{(0)}_G&{\rm if~} G{\rm ~is~topologically~trivial} \\
R^{(1)}_G&{\rm if~} G{\rm ~is~topologically~nontrivial}
\end{array}
\right.
\ee
Then the following theorem holds.


\FIGURE[ht]{
\includegraphics[width=4.9truecm]{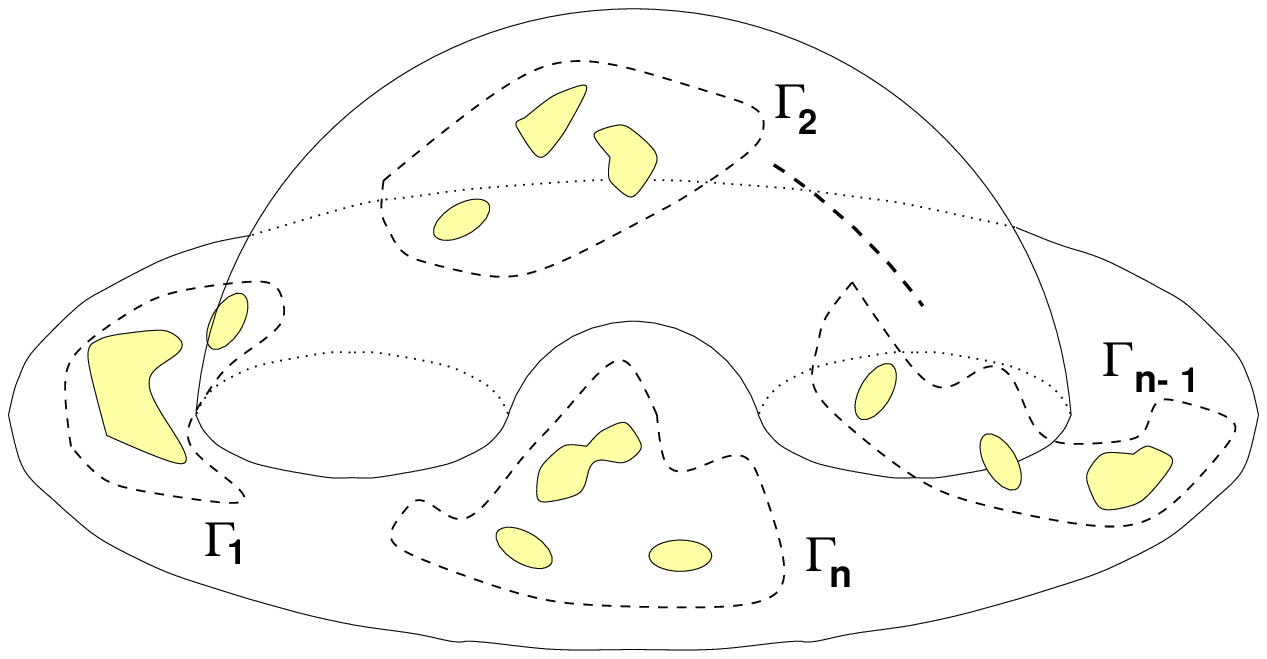}
\includegraphics[width=4.9truecm]{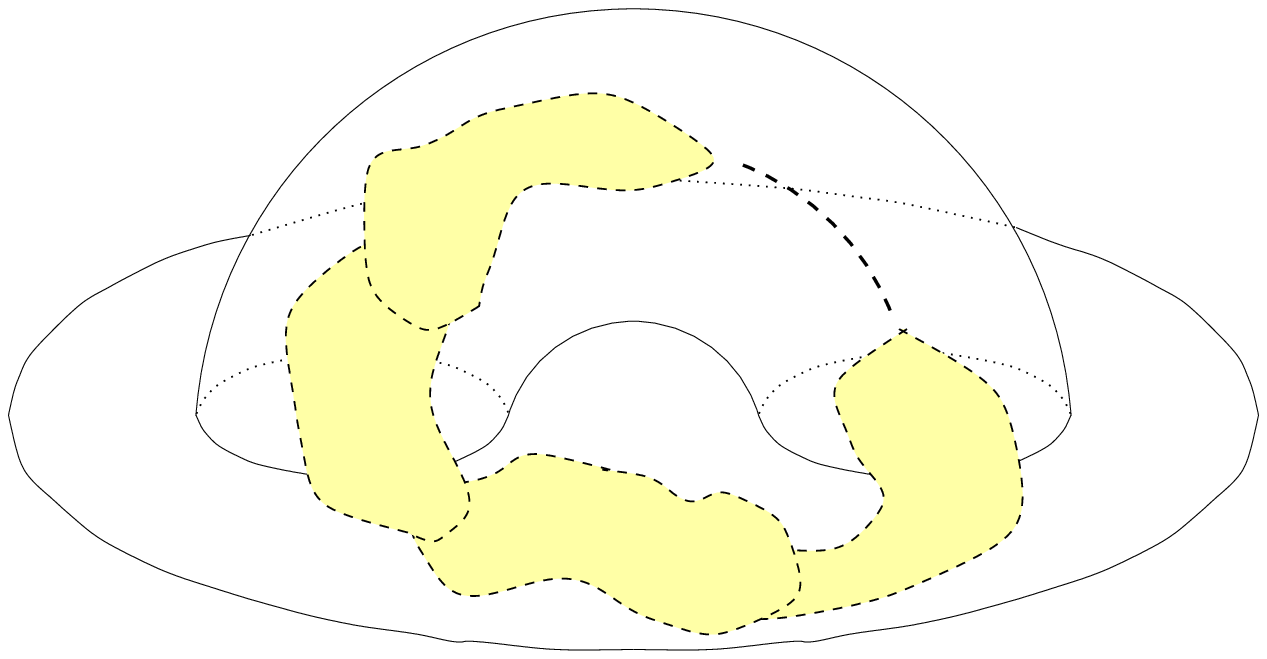}
\includegraphics[width=4.9truecm]{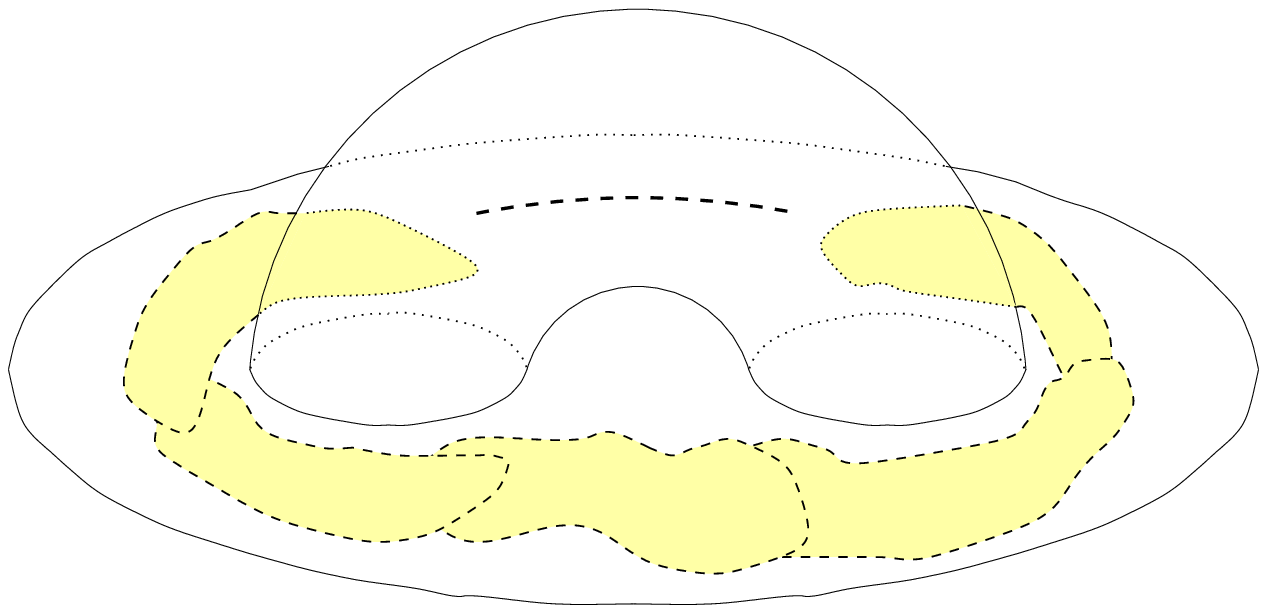}
{\vspace{-2truemm}}
\centerline{~~~~~~~~~~~~~~~~~{\bf (a)}~~~~~~~~~~~~~~~~~~~~~~~~~~~~~~~~~{\bf (b)}~~~~~~~~~~~~~~~~~~~~~~~~~~~~~~~~~~{\bf (c)}~~~~~~~~~~~~~~}
{\vspace{-8truemm}}
\caption{Illustration for Theorem 3.}
}

\noindent
{\bf Theorem 3.} {\it If the graph $G$ belongs to the class $\Omega_d$,
then $R_G$  leads to a finite integral.}

\noindent
{\bf Proof.}  If $G$ is planar, then $R_G=R_G^{(0)}$.  It is known
that the planar version of Bogoliubov-Parasuik's formula renders
all divergent planar diagrams finite\cite{rivass}. Thus we consider
 nonplanar graph $G$. In this case $R_G=R_G^{(1)}$. Let us
draw $G$ on a surface $\Sigma_g$. 
The idea of the proof that $R_G$ leads to a finite 
integral is simple. We just have to show that $R_G$ satisfies the
conditions of Theorem 2. 
There are three potentially distinct cases to consider:
\begin{enumerate}
\item  $G$ has a disjoint set of topologically trivial proper 1PI
subgraphs $\{ \Gamma_1,\Gamma_2,\ldots,\Gamma_n\}$ such that
each renormalization part $\gamma\in \Re(G)$ is contained in
one of them (see figure 14(a)).
\item $G$ has overlapping renormalization parts forming a subgraph 
$\gamma_1$ with $\omega(\gamma_1)\ge 0$, 
which wraps a homologically nontrivial cycle of $\Sigma_g$ 
(see figure 14(b)).
\item $G$ has overlapping renormalization parts forming a
subgraph $\gamma_2$ with  $\omega(\gamma_2)\ge 0$, which
wraps a topologically nontrivial, but homologically trivial cycle
of $\Sigma_g$ (see figure 14(c)). 
\end{enumerate}
Let us analyze these cases:

\noindent
Case 1. 

\noindent
For each $\Gamma_i$ in figure 14(a), the subtracted integrand
$R^{(0)}_{\Gamma_i}$ satisfies the condition $\omega <0$. 
Let $\Gamma$ be a subgraph of $G$ which wraps a homologically
nontrivial cycle of $\Sigma_g$. Let 
$\Gamma_1,\ldots,\Gamma_k \subset \Gamma$ be a disjoint set of
topologically trivial 1PI subgraphs of $\Gamma$.
Using the relation
$$
\omega(\Gamma)=\omega(\Gamma/\{ \Gamma_1,\ldots,\Gamma_k\})+
\sum_{a=1}^k \omega(\Gamma_a)
$$
and the fact that $\omega(\Gamma)<d$,
$
\omega(R^{(0)}_{\Gamma_i})<0
$,
$i=1,\ldots,k$,
we find that
$$
\omega(R_{\Gamma})<d
$$
One can similarly show that all other subtracted subgraphs satisfy the
conditions of Theorem 2.   


\noindent
Cases 2 and 3.

\noindent
If $G$ were not in the class $\Omega_d$, 
for the subtraction of divergences of $\gamma_1$ and $\gamma_2$ 
one would have to introduce nonplanar counterterms (geometrically
it means that we pinch  the handles of $\Sigma_g$). We do
not know how to deal with this situation. 
But since $G$ is assumed to be in the class $\Omega_d$, we
do not have to subtract the graphs $\gamma_1$ and $\gamma_2$ as a
whole.  Thus the global structure of the graph $G$ remains unchanged
as a result of the subtraction procedure. 
The argument given in the Case 1 then applies here as well.~~~  {\bf q.e.d.}

\subsection{Generation of subtractions by counterterms}
For a given scalar  NQFT in $d$ dimensions, 
we have seen how to renormalize an individual noncommutative Feynman
graph $G$ from the class $\Omega_d$ 
by applying the recursion formula to the integrand $\I_G$. 

If scalar NQFT is not renormalizable in
the commutative limit, then the class $\Omega_d$ is smaller than 
the class of all diagrams of the theory. 
In a commutative QFT it is
possible to renormalize a nonrenormalizable theory by including
counterterms with an arbitrarily large number of powers of momentum and 
with an arbitrarily large number of external lines. 
Our subtraction 
procedure works only for the graphs from the class $\Omega_d$.  
Thus if NQFT is not renormalizable in the commutative limit, then
it is not possible to renormalize an arbitrary graph by
the introduction of counterterms in the action of the
form 
$$\tr ~L_{ct} (\Phi({\hat x}),{\hat \partial}\Phi({\hat x}), 
{\hat \partial}{\hat \partial}\Phi({\hat x}),\ldots)
$$
where
${\hat \partial}_{\mu}\equiv \theta_{\mu\nu}^{-1}[{\hat x}^{\nu},\cdot]$ is
the noncommutative analog of the derivative. 

Unfortunately, even for the scalar field theories which 
are renormalizable in the commutative limit $\theta=0$ 
(e.g. $\phi\star\phi\star\phi$ theory in six dimensions) the
class $\Omega_d$ is smaller than the class of all diagrams of
the theory. \footnote{This was pointed out to us by Shiraz Minwalla and
Mark Van Raamsdonk. The counter-example is given in figure 15. 
In the original version of this paper, we stated that 
if NQFT is renormalizable in the commutative limit, then 
the class $\Omega_d$ is {\it equal} to the class of all diagrams
of the theory. This wrong statement led us to conclude that 
$\phi\star\phi\star\phi$ theory in six dimensions is renormalizable.}
In ref.\cite{minwalla}, the diagram in 
figure 15(a) 
is shown to be divergent 
in six dimensions for $n\ge 3$.  An easy way to see this is
to note that $P_2$ for this graph is equal to $P_0$ for the graph in 
figure 15(b)
(see section 2.2 for the definition of $P_0$ and $P_2$)
The graph in 
figure 15(b) 
is divergent in six dimensions for $n\ge 3$.

The other way to see the divergence of the graph in figure 15 in
six dimensions is  
to note that the disjoint subgraph $\gamma$ formed by the lines
$1,2,\ldots,2n-1,2n$ has $\omega(\gamma)=6n-4n=2n$ and $c(\gamma)=1$.
It means that the condition (1) of Theorem 2 is violated if $n\ge 3$:
$\omega(\gamma)-6 \ge 0$.

In general the graphs of the type shown in 
figure 16 
are not in the class $\Omega_6$ for the $\phi\star\phi\star\phi$
theory. The reason is the following. 
Each of the subgraphs $\gamma_i$ in figure 16 has two
external lines and thus  $\omega(\gamma_i)=2$. But 
$c(\gamma_1 \cup \gamma_2 \cup \cdots \cup \gamma_n)=1$. 

\FIGURE[ht]{\hbox{\raise25pt\hbox{
\includegraphics[width=7.5truecm]{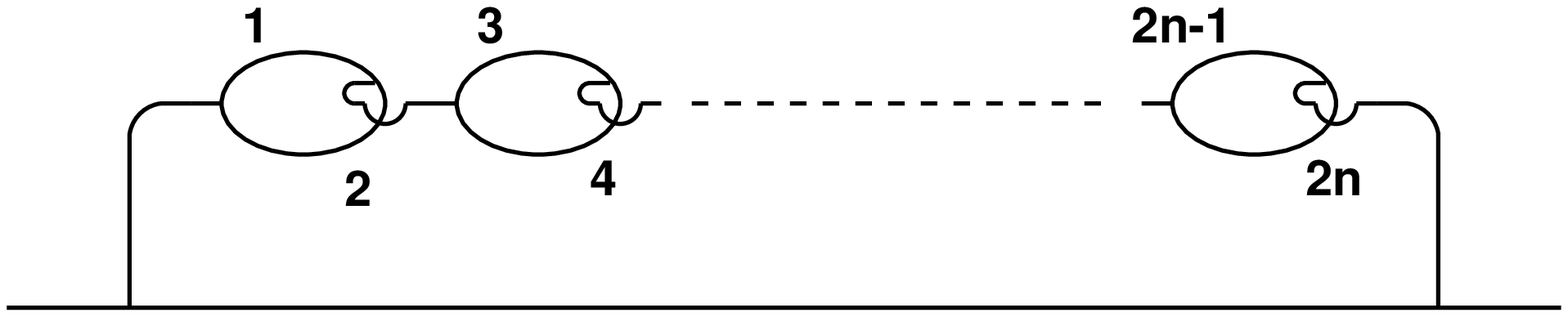}}}
\hspace{3truemm}
\hbox{
\includegraphics[width=4truecm]{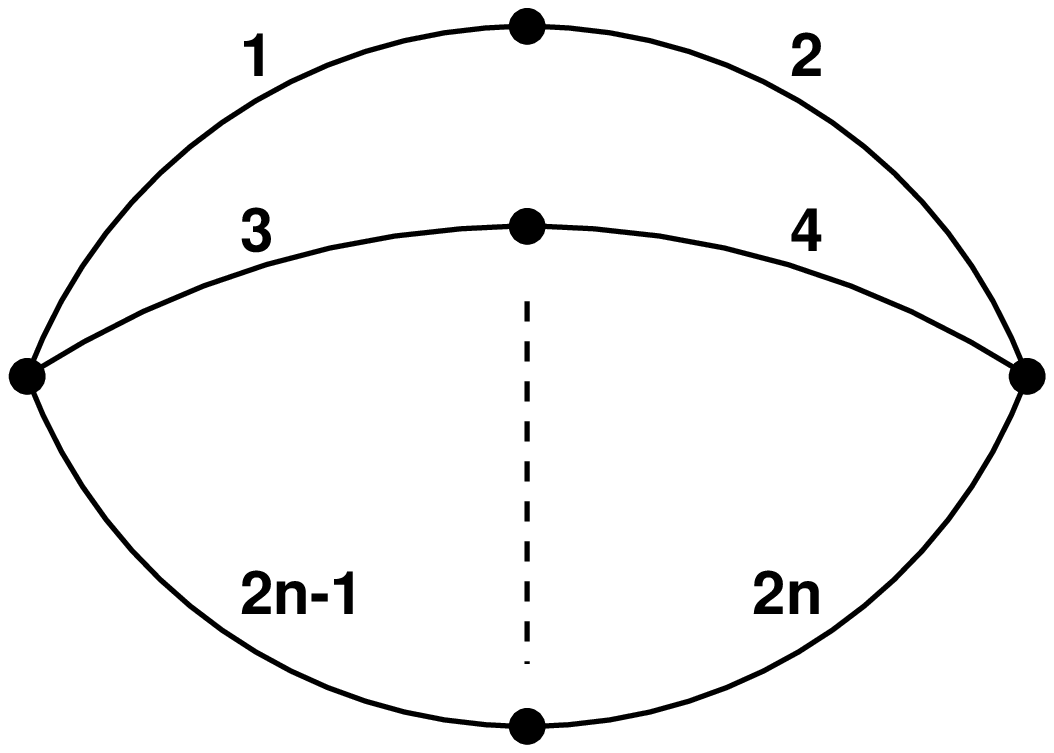}}
\centerline{~~~~~~~~~~~{\bf (a)}~Minwalla-Raamsdonk-Seiberg's~~~~~~~~~~{\bf (b)}~Representation of $P_2$~~~~~}
\centerline{counter-example.~~~~~~~~~~~~~~~~~~~~~~~~~~~~~~~~~~~~~~~~~~~~~~}
\vspace{-8truemm}
\caption{}
}


Let us show, on the example of  $\phi\star\phi\star\phi$ theory
in six dimensions,  that the recursive subtraction 
procedure of section 4.1 is equivalent to the counterterm approach. 
Although the subtraction procedure of section 4.1 is incapable of
removing all divergences of the theory, the analysis given in this
section is
useful for the discussion about Wess-Zumino model given in section 5.
The following discussion is 
completely parallel to the one given in ref.\cite{collins} for the 
commutative QFT.   
Let us decompose the Lagrangian as follows:
\be
{\cal L}={\cal L}_0+{\cal L}_b+{\cal L}_{ct}
\la{lagran}
\ee  
Here $\l_0$ is the free Lagrangian 
\be
\l_0={1\ov 2} (\partial \phi)^2+{m^2\ov 2} \phi^2
\ee
with $m$ being the renormalized mass. The rest of the Lagrangian, 
$\l_I=\l_b +\l_{ct}$,is the interaction, and consists of two terms.
The first, which we will call the basic interaction, is
\be
\l_b=(g/3) (\phi\star\phi\star\phi)
\ee

\FIGURE{
\includegraphics[width=10truecm]{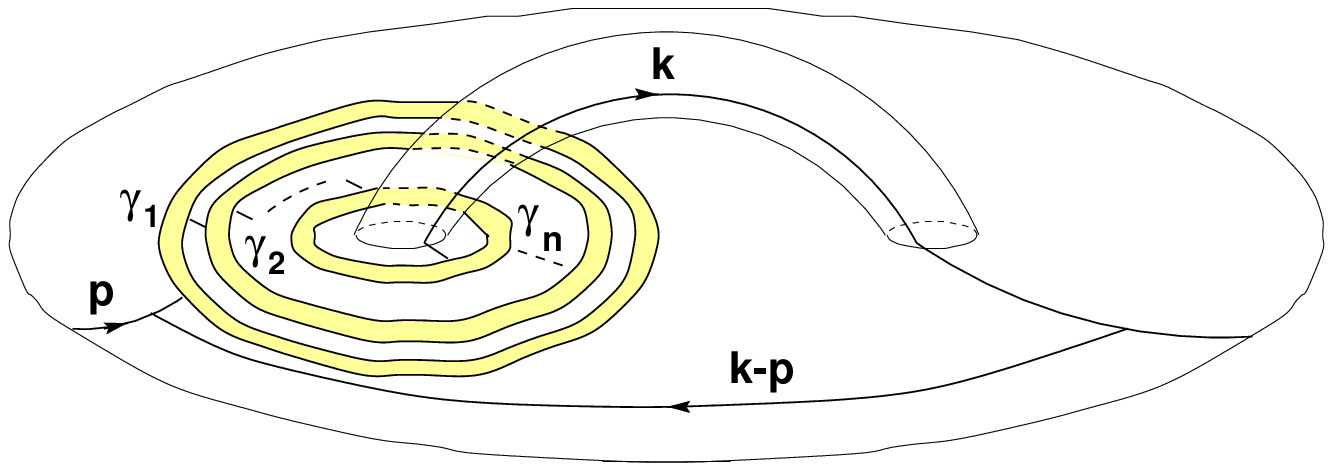}
\caption{A general diagram not included in class $\Omega_6$.}
}

\FIGURE[ht]{
\includegraphics[width=6truecm]{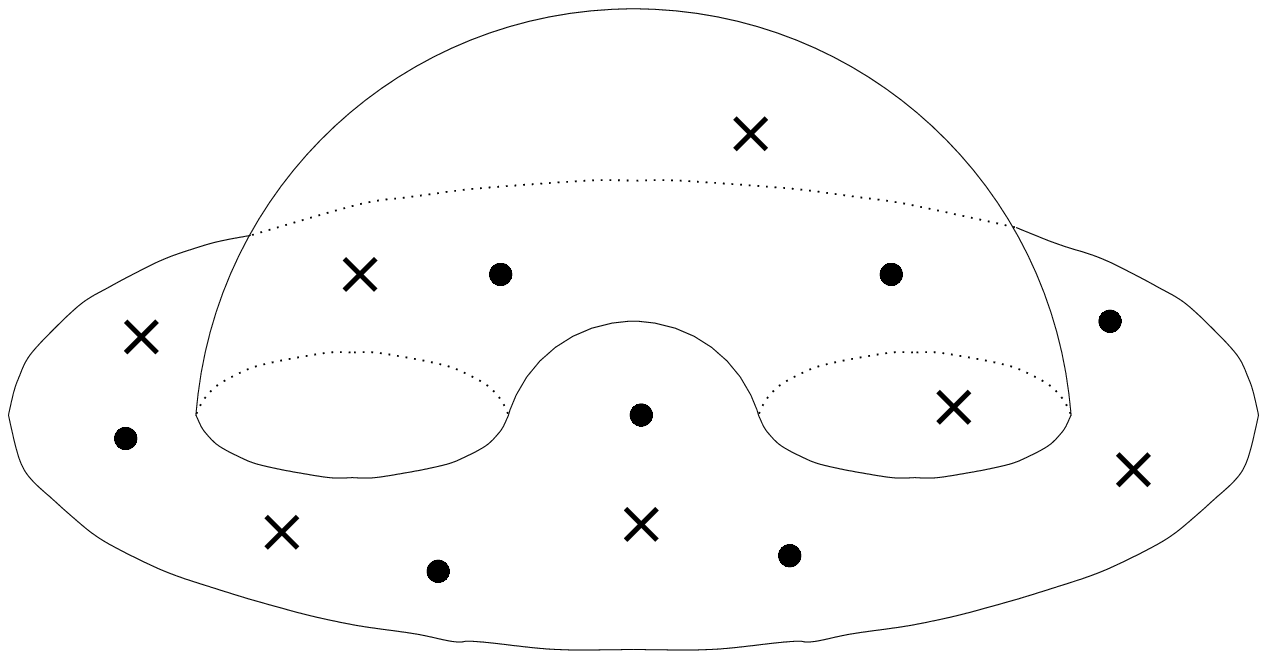}
\hspace{2truecm}
\includegraphics[width=6truecm]{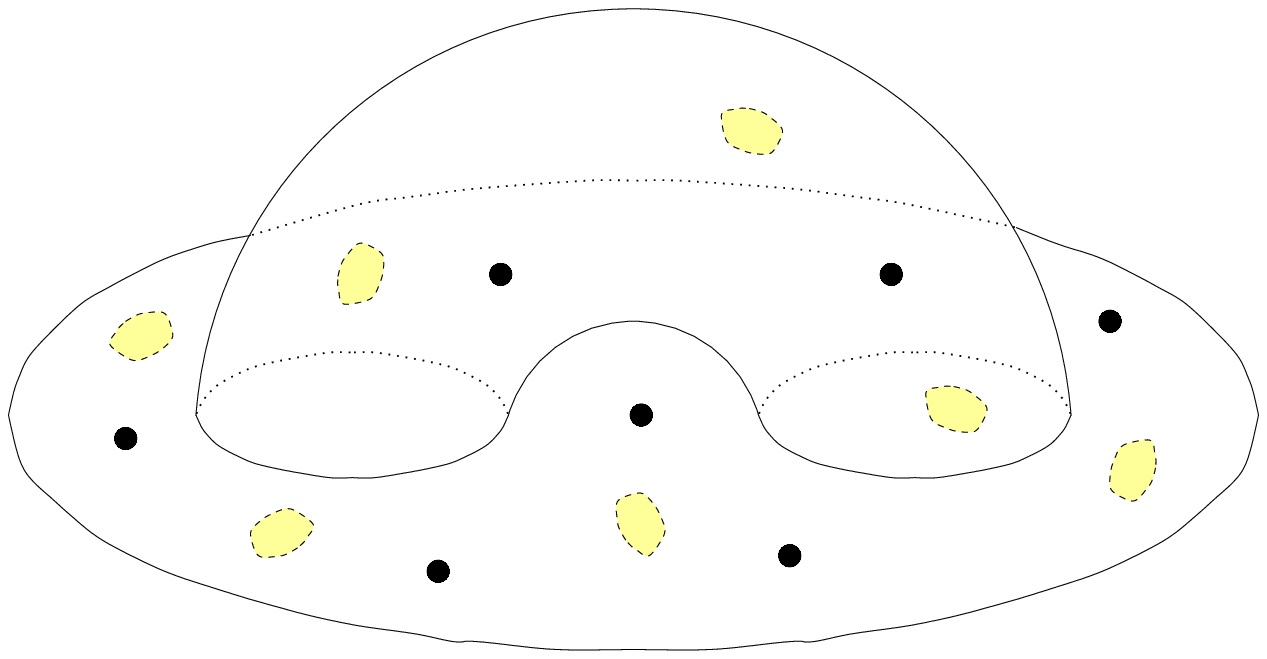}
{\vspace{2truemm}}
\centerline{~~~~~~~~~~~~{\bf (a)}~~~~~~~~~~~~~~~~~~~~~~~~~~~~~~~~~~~~~~~~~~~~~~~~~~{\bf (b)}~~~~~~~~~~~~~~~~}
{\vspace{-8truemm}}
\caption{The  counterterm and basic graphs. $\times$ and $\bullet$ stand for the counterterm
and basic vertices respectively.}
}

The second term is the counterterm Lagrangian and it is defined 
as follows. Let $C_2(G,k_1,k_2)$ be the overall counterterm for the 
planar 1PI graph $G$ with two external lines. Let $C_3(G,k_1,k_2,k_3)$
be the overall counterterm for the planar 1PI graph $G$ with three
external lines. Note that $C_2$ and $C_3$  do not
contain the phase factors associated with the external momenta.
Then the counterterm action $S_{ct}$ reads 
\bea
S_{ct}&=&\sum_{2-{\rm point~}G}{1\ov 2}\int d^6k_1 d^6k_2 d^6k_3
\phi(k_1)C_2(G,k_1,k_2)\phi(k_2) \non
&&~~+
\sum_{3-{\rm point~}G}{1\ov 3}\int d^6k_1 d^6k_2 d^6k_3  
\phi(k_1)\phi(k_2)\phi(k_3) C_3(G,k_1,k_2,k_3) \times \non
&&~~~~~~~~~~~~~~~~~\times \ex\LB i\theta_{\mu\nu}
(k_1^{\mu}k_2^{\nu}+k_1^{\mu}k_3^{\nu}+k_2^{\mu}k_3^{\nu})
\RB
\eea
Thus the counterterm Lagrangian is 
\be
\l_{ct}=\delta Z (\partial\phi)^2/2 +\delta m^2\phi^2/2+\delta 
g (\phi\star\phi\star\phi)/3
\la{cterm}
\ee
with
\bea
\delta Z&=&\sum_{2-{\rm point~}G} [{\rm Coefficient~of~}-p^2{\rm ~in~}
C_2(G)]\non
\delta m^2&=&\sum_{2-{\rm point~}G} [{\rm Coefficient~of~}p^0{\rm ~in~}
C_2(G)]\non   
\delta g&=&\sum_{3-{\rm point~}G} C_3(G)
\la{vertices}
\eea

Consider the full $N$-point Green's function $G_N$ at order $g^L$ in the
NQFT with the Lagrangian given by \eq{lagran}. The term of order $g^L$
in the perturbation expansion of $G_N$ has vertices generated by the
different terms in $\l_b+\l_{ct}$. There will be graphs with all of their 
vertices being the basic interaction $\l_b$. The other graphs will
contain one or more of the counterterm vertices generated by $\l_{ct}$
\eq{cterm}. A generic diagram which contains counterterm vertices looks 
like the one shown in figure 17(a). If we replace 
each counterterm vertex in the graph in figure 17(a)
 by the  sum over  overall-divergent 1PI graphs as in \eq{vertices},
then each term in the resulting multiple sum corresponds to
a unique basic graph as  in figure 17(b). 
On the other hand,
according to the subtraction formula \eq{subtract}, for a
 graph $G$ of genus $g$ we subtract all possible  
 disjoint unions of divergent topologically trivial (planar) 1PI subgraphs. 
The analysis is completely parallel to the one in commutative
 case\cite{collins},
with  a simplification  due to the combinatorics in the noncommutative
case. The point is that
all ribbon graphs come with the combinatorial factor 1. Thus 
the recursive subtraction procedure of section 4.1 is 
equivalent to the counterterm approach.

\section{Conclusions and discussions}
We proved the convergence of some classes of diagrams
in massive scalar quantum field theories on noncommutative
$\re^d$  and formulated a general convergence theorem for
the noncommutative Feynman graphs. Although we did not
prove the  convergence theorem  in its general form, we made 
it very plausible by demonstrating its universal character.
We should also mention that we analyzed numerous other examples
not discussed in this paper and found that they are in complete
agreement with the statements of the general convergence theorem.

We proposed a recursive
subtraction formula for divergent Feynman graphs and showed that 
for the graphs in 
class $\Omega_d$ it leads to finite integrals. 
For a generic
 scalar noncommutative quantum field theory 
on $\re^d$, the class $\Omega_d$ is  smaller than the class of all 
diagrams in the theory. 
This leaves open the  question of perturbative renormalizability
of noncommutative field theories.  
As explained in section 4.2,
the problematic graphs (the graphs that are not in the class $\Omega_d$)
are of the type shown in figure 16. All the rings 
$\gamma_1,\ldots,\gamma_n$ wrap a single cycle, but each ring 
has $\omega > 0$.  For a large enough number of rings, the
subgraph formed by their disjoint union  
will not 
satisfy  the condition (1) of Theorem 2, i.e. the graph will
diverge. A natural way to avoid the violation
of condition (1) in Theorem 2 by the accumulation
of positive $\omega$'s  is to enforce  the condition $\omega \le 0$ for
the subgraphs.
This situation is realized in supersymmetric 
theories.\footnote{The non-supersymmetric 
$\phi\star\phi\star\phi$ theory in four dimensions also has only logarithmic
divergences, but it is a trivial theory.}.
As
an example consider a noncommutative version of
 supersymmetric Wess-Zumino model in
four dimensions:  
$$
S[\Phi]=\int d^4x d^2\theta d^2{\bar\theta}^2 ~~\Phi^+\star\Phi +\left\{
\int d^4x d^2\theta \left[  m\Phi^2 + g~\Phi\star\Phi\star\Phi\right]
+~{\rm h.c.}~\right\}
$$
It is well known that commutative Wess-Zumino model has only
logarithmic divergences.  Thus it is plausible that the noncommutative
Wess-Zumino model is renormalizable.

\bigskip

\noindent
{\bf Acknowledgments}
We are grateful to Y.M.~Makeenko, H.~Nastase, M.~Ro\v{c}ek and
G.~Sterman
for  useful discussions and suggestions. We would also like
to thank S.~Minwalla and M.V.~Raamsdonk for pointing out a
 mistake regarding the class $\Omega_d$ in the original version 
of this paper.
This work was supported in part by NSF grant PHY-9722101.

\bigskip

\setcounter{section}{0}
\setcounter{subsection}{0}

\appendix

\section{Parametric integral representation}
Using the integral representations for the propagators and the
$\delta^d$ function in \eq{IG}\footnote{See ref.\cite{itzyk} } one finds
\bea
&&\!\!\!\!\int_0^{\infty} \prod_{l=1}^I d\a_l \int \prod_{v=1}^V d^dy_v
{\rm e}^{-\sum_l \a_l m_l^2} \int\prod_l d^dk_l \non
&& \ex \left[ -\sum_l a_lk_l^2-
i\sum_v y_v\cdot (P_v-\sum_l\e_{vl} k_l)\right. \non 
&&~~~~+\left. i (\sum_{m,n} I_{mn} 
\theta_{\mu\nu}k_m^{\mu} 
k_n^{\nu}+
\sum_{m,v} J_{mv} \theta_{\mu\nu}k_m^{\mu} P_v^{\nu} )\right]
\la{aone}
\eea
Integration over the momenta $k$ in \eq{aone} gives
\bea
I_G(P)&=&\pi^{{Id\ov 2}}\int_0^{\infty}\prod_l d\a_l 
{\rm e}^{-\sum_l \a_l m_l^2}\int\prod_v d^dy_v {\rm e}^{-i\sum_v y_v\cdot P_v}
 ({\rm det}\A)^{-{1\ov 2}} \non
&& \ex\left\{-{1\ov 4}[(J\eta)^{\mu}_m+(y\e)^{\mu}_m ]
(\A^{-1})^{\mu\nu}_{mn} [(J\eta)^{\nu}_n +(y\e)^{\nu}_n]\right\}
\eea
Making the following change of integration variables
\bea
y_1&=&z_1+z_V \non
y_2&=&z_2+z_V \non
   &\vdots&    \non
y_{V-1}&=&z_{V-1}+z_V  \non
y_v&=&z_V
\eea
the jacobian of which is one, and using the fact that $\sum_v \e_{vl}=0$, 
one finds
\bea
I_G(P)&=&\pi^{{Id\ov 2}}(2\pi)^d \delta^{(d)}(\sum_v P_v)\int_0^{\infty}
\prod_{l=1}^I d\a_l  {\rm e}^{-\sum_l \a_l m_l^2} ({\rm det}\A)^{-{1\ov 2}}
\int \prod_{v=1}^{V-1} d^d z_v {\rm e}^{-i\sum_{v=1}^{V-1} z_v\cdot P_v} \non
&&\ex\left\{ -{1\ov 4}[(J\eta)^{\mu}_m+(z\e_V)^{\mu}_m ]
(\A^{-1})^{\mu\nu}_{mn} [(J\eta)^{\nu}_n +(z\e_V)^{\nu}_n]\right\}
\eea
where $(z\e_V)^{\mu}_m\equiv \sum_{v=1}^{V-1} z_v^{\mu}\e_{vm}$.
Performing the z-integrals, one finds \eq{alpha}.

\section{Proof of the lemma}
For simplicity we prove the lemma for the $n=3$ case. Generalization
to the case of arbitrary $n$ is straightforward. 
 Let us consider a graph $\pic{4}{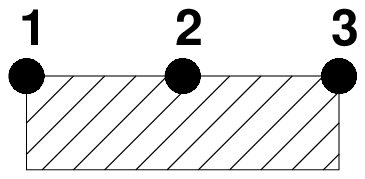}$  with  three of its vertices
labeled. 
We will denote by crosses vertices that are identified.

For $n=3$ \eq{identity} reads
\bea
P_0(\pic{4}{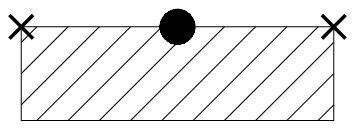}) &=& P_0(\pic{4}{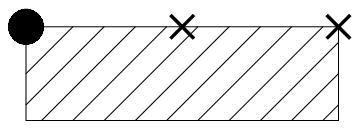})\oplus 
P_0(\pic{4}{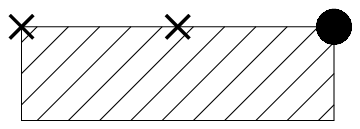})\,\,\,\,{\rm or\,\,\,\,equivalently}\nonumber\\
P_0(\pic{4}{i1.ps})&\oplus& P_0(\pic{4}{i2.ps})\oplus 
P_0(\pic{4}{i3.ps})=0\,\,\,\, .
\la{nthree}
\eea

\noindent
The set of trees of the graph 
$\pic{4}{i2.ps}$
can be written as 
the union of 
two sets: the set of trees of that directly link vertices $1$ and $2$, 
and the  set of trees of   that directly link vertices $1$ and $3$
\be
{\cal T}(\pic{4}{i2.ps})={\cal T}(\pic{12}{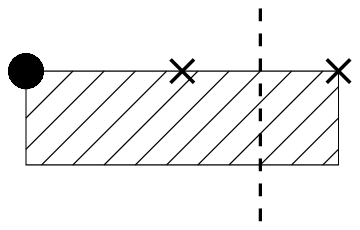})\cup 
{\cal T}(\pic{4}{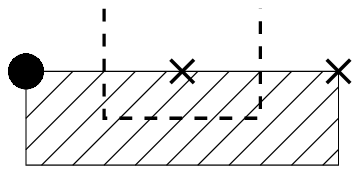})
\ee
Example: 
\bea
{\cal T}(\pic{10}{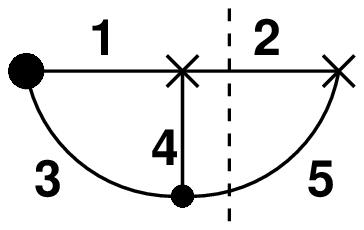})&=&
{\Big\{}\,\, \pic{6}{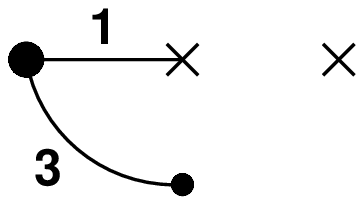}\,\,,\,\, \pic{6}{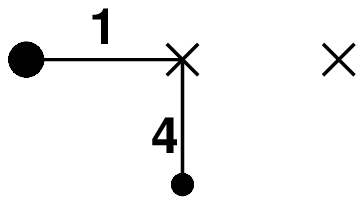}\,\,,
\,\, \pic{6}{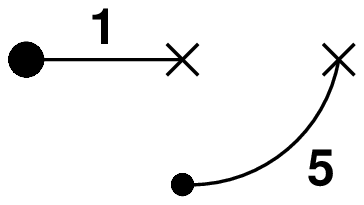}\,\,,\,\, \pic{6}{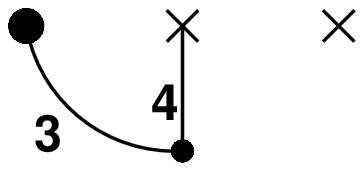}\,\,{\Big\}}=\nonumber\\
&=&
{\Big\{}\,\, \pic{6}{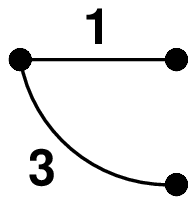} ,\,\, \pic{6}{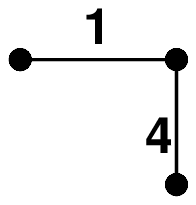}\!,
\! \pic{6}{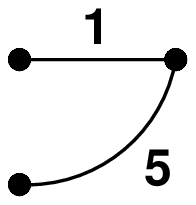}\,\,,\,\, \pic{6}{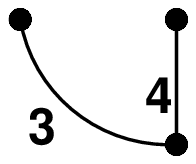}{\Big\}}\\
{\cal T}(\,\pic{6}{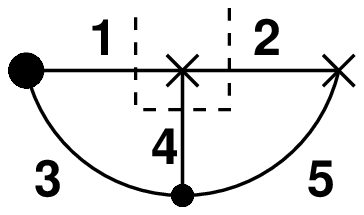}\,)&=&\Big\{\,\,\pic{6}{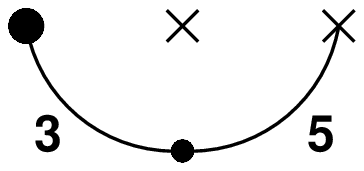}\,\,
\Big\}=
\Big\{\,\,\pic{6}{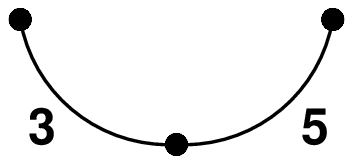}\,\,\Big\}\nonumber
\eea

Similarly, we can split 
the trees of 
$\pic{4}{i3.ps}$
 and  
$\pic{4}{i1.ps}$
as follows:
\be
{\cal T}(\pic{4}{i3.ps})={\cal T}(\pic{12}{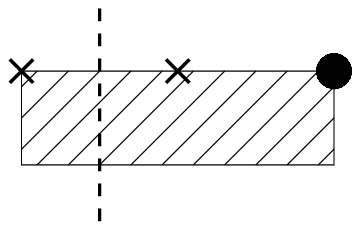})\cup 
{\cal T}(\pic{4}{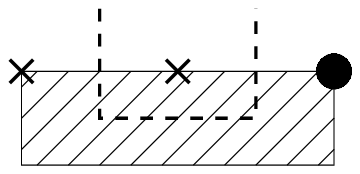})
\ee
\be
{\cal T}(\pic{4}{i1.ps})={\cal T}(\pic{12}{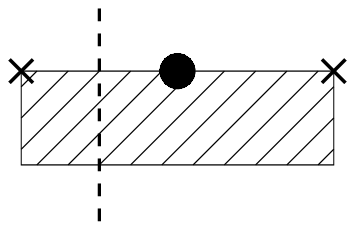})\cup 
{\cal T}(\pic{12}{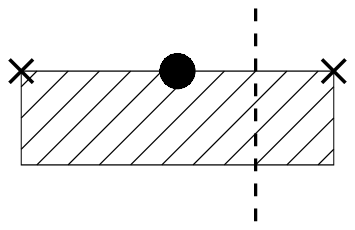})
\ee
We notice that the chords 
associated with 
$\pic{4}{i21.ps}$
are identical to the chords associated with
 $\pic{4}{i31.ps}$ and therefore cancel in $\oplus$ sum.
Using the relations
\be
{\cal T}^*(\pic{12}{i32.ps})={\cal T}^*(\pic{12}{i11.ps})
\ee
and
\be
{\cal T}^*(\pic{12}{i22.ps})={\cal T}^*(\pic{12}{i12.ps})
\ee
where ${\cal T}^*$ denotes the chord set,
it is easy to verify that \eq{nthree} holds.

\end{document}